\documentclass[12pt,preprint]{aastex}

\newcommand{\kms}{km s$^{-1}$}
\newcommand{\Bband}{B}
\newcommand{\Vband}{V}
\newcommand{\BminusV}{({\Bband}{\rm -}{\Vband})}
\newcommand{\bminusv}{[{\Bband}{\rm -}{\Vband}]}

\newcommand{\EBV}{E\bminusv}
\newcommand{\Ebv}{E\BminusV}

\newcommand{\ebv}{$E\BminusV$}
\newcommand{\halpha}{H$\alpha$}

\newcommand{\hbeta}{H$\beta$}
\newcommand{\hgamma}{H$\gamma$}
\newcommand{\hdelta}{H$\delta$}
		
\newcommand{\bvri}{\protect\hbox{$BV\!RI$}}		
\newcommand{\bvi}{\protect\hbox{$BV\!I$}}		
\newcommand{\vri}{\protect\hbox{$V\!RI$}}		
		
\newcommand{\ssp}{\def\baselinestretch{1.0}\large\normalsize}
\newcommand{\gtrsi}{\mathrel{\hbox{\rlap{\hbox{\lower4pt\hbox{$\sim$}}}\hbox{$>$}}}}
\newcommand{\vi}{\mbox{$V\!-\!I$}}

\shorttitle{Study of SN 1999gi} \shortauthors{Leonard et al.}

\begin{document}

\title{A Study of the Type II-Plateau Supernova 1999gi, and the Distance to its
Host Galaxy, NGC 3184}

\vspace{2cm}

\author{Douglas C. Leonard\altaffilmark{1}, 
Alexei V. Filippenko\altaffilmark{2}, 
Weidong Li\altaffilmark{2}, 
Thomas Matheson\altaffilmark{3}, 
Robert P. Kirshner\altaffilmark{3},
Ryan Chornock\altaffilmark{2}, 
Schuyler D. Van Dyk\altaffilmark{4},
Perry Berlind\altaffilmark{3},
Michael L. Calkins\altaffilmark{3}, 
Peter M. Challis\altaffilmark{3},
Peter M. Garnavich\altaffilmark{3,5},
Saurabh Jha\altaffilmark{3,2}, and
Andisheh Mahdavi\altaffilmark{3}
}

\altaffiltext{1}{Department of Astronomy, University of Massachusetts, Amherst,
MA 01003-9305; leonard@nova.astro.umass.edu}

\altaffiltext{2}{Department of Astronomy, University of California, Berkeley,
CA 94720-3411}

\altaffiltext{3}{Harvard-Smithsonian Center for Astrophysics, 60 Garden St.,
Cambridge, MA 02138}

\altaffiltext{4}{Infrared Processing and Analysis Center, 100-22, California
Institute of Technology, Pasadena, CA 91125}

\altaffiltext{5}{Present address: Physics Department, University of Notre Dame,
Notre Dame, IN 46556}

\vspace{1cm}

\begin{abstract}

We present optical spectra and photometry sampling the first six months after
discovery of supernova (SN) 1999gi in NGC~3184.  SN~1999gi is shown to be a
Type II-plateau event with a photometric plateau lasting until about 100 days
after discovery.  The reddening values resulting from five independent
techniques are all consistent with an upper bound of $\Ebv < 0.45$ mag
established by comparing the early-time color of SN~1999gi with that of an
infinitely hot blackbody, and yield a probable reddening of $\Ebv = 0.21 \pm
0.09$ mag.  Using the expanding photosphere method (EPM), we derive a distance
to SN~1999gi of $11.1^{+2.0}_{-1.8}$ Mpc and an explosion date of 1999 December
$5.8^{+3.0}_{-3.1}$, or $4.1^{+3.0}_{-3.1}$ days prior to discovery.  This
distance is consistent with a recent Tully-Fisher distance derived to NGC~3184
($D \approx 11.59$ Mpc), but is somewhat closer than the Cepheid distances
derived to two galaxies that have generally been assumed to be members of a
small group containing NGC~3184 (NGC 3319, $D = 13.30 \pm 0.55$ Mpc, and NGC
3198, $D = 13.80 \pm 0.51$ Mpc).

We reconsider the upper mass limit ($9^{+3}_{-2} {\rm \ M}_{\odot}$) recently
placed on the progenitor star of SN~1999gi by Smartt et al. (2001, 2002) in
light of these results.  Following the same procedures, but using the new data
presented here, we arrive at a less restrictive upper mass limit of
$15^{+5}_{-3} {\rm \ M}_{\odot}$ for the progenitor.  The increased upper limit
results mainly from the larger distance derived through the EPM than was
assumed by the Smartt et al. analyses, which relied on less precise (and less
recent) distance measurements to NGC~3184.

Finally, we confirm the existence of ``complicated'' P-Cygni line profiles in
early-time and later photospheric-phase spectra of SN~1999gi.  These features,
first identified by Baron et al. (2000) in spectra of SN~1999em as
high-velocity absorptions in addition to the ``normal'' lower-velocity
component, are here verified to be true P-Cygni profiles consisting of both an
absorption trough and an emission peak at early times.  In the earliest
spectrum, taken less than a day after discovery, the features extend out to
nearly $-30,000$ \kms, indicating the existence of very high-velocity material
in the outer envelope of SN~1999gi.

\end{abstract}

\medskip

\section{INTRODUCTION}
\label{sec:introduction}

Supernova (SN) 1999gi was discovered by Nakano et al. (1999) on 1999 December
9.82 (UT dates are used throughout this paper) at an unfiltered magnitude of $m
\approx 14.5$ in the nearly face-on ($i < 24^{\circ}$, from the Lyon-Meudon
Extragalactic Database [LEDA\footnote{\url{http://leda.univ-lyon1.fr}}]) SBc
galaxy NGC~3184.  The identification of hydrogen in an early-time spectrum
quickly defined it as a Type II event (Nakano et al. 1999; see Filippenko 1997
for a review of SN types), and the absence of the SN on CCD images of the same
field taken 6.64 and 7.32 days earlier (Trondal et al. 1999, limiting
unfiltered magnitude 18.5, and Nakano et al. 1999, limiting unfiltered
magnitude 19.0, respectively) implies that it was discovered shortly after
explosion.

There have been two previous investigations of SN~1999gi.  In the first,
Leonard \& Filippenko (2001) examine a single epoch of optical
spectropolarimetry of SN~1999gi taken 107 days after discovery. They find an
extraordinarily high degree of linear polarization, $p_{\rm max} = 5.8\%$,
where $p_{\rm max}$ is the highest level of polarization observed in the
optical bandpass.  If intrinsic to SN~1999gi, such polarization implies an
enormous departure from spherical symmetry (H\"{o}flich 1991).  However,
Leonard \& Filippenko (2001) conclude that the majority of the polarization is
likely due to interstellar dust, and is not intrinsic to the SN.  From
photometry reported in various IAU Circulars, Leonard \& Filippenko (2001)
tentatively classify SN~1999gi as a Type II-plateau supernova (SN II-P).  In
addition, the total flux spectrum of SN~1999gi from day 107 and a comparable 
spectral epoch of SN~1999em, a classic SN II-P (Leonard et al. 2002 [hereafter
L02]; Hamuy et al. 2001), show great spectral similarity, suggesting that
they may have been quite similar events.

In the second study, Smartt et al. (2001; hereafter S01) examine pre-explosion
archival {\it Hubble Space Telescope (HST)} images of NGC~3184 and use the lack
of a progenitor-star detection in the pre-discovery frame, along with an
estimated distance of $D = 7.9$ Mpc, to set an upper limit on the absolute
magnitude of the progenitor for SN~1999gi.  This is then translated into an
upper mass limit of $9^{+3}_{-2}\ {\rm M}_{\odot}$ for the progenitor of
SN~1999gi through comparison with stellar evolution models.  A subsequent
reanalysis of the same data by Smartt et al. (2002; hereafter S02), using
improved models, confirms this limit.  Since stars with initial mass $\lesssim
8\ {\rm M}_{\odot}$ are not expected to undergo core collapse (e.g., Woosley \&
Weaver 1986, and references therein), this upper bound sets very tight
constraints on the possible mass of the progenitor, a fact that has important
implications for the nature of the progenitors of SNe II-P. Indeed, other than
the Smartt et al. studies, progenitor masses for SNe II-P are virtually
unconstrained by direct observation.\footnote{Mass constraints on the
progenitors of other types of core-collapse SNe have been obtained through
studies of their environments; see, e.g., Van Dyk et al. (1999b).}  Although
the progenitors of SN 1961V (Goodrich et al. 1989; Filippenko et al. 1995; Van
Dyk, Filippenko, \& Li 2002), SN 1978K (Ryder et al. 1993), SN 1987A (e.g.,
White \& Malin 1987; Walborn et al. 1987), SN 1993J (Filippenko 1993; Aldering,
Humphreys, \& Richmond 1994; Cohen, Darling, \& Porter 1995), and SN 1997bs
(Van Dyk et al. 1999a) have been identified, all of these SNe II were peculiar.

In this paper, we present 15 optical spectra and 30 photometric epochs of
SN~1999gi sampling the first 169 and 174 days since its discovery,
respectively, and derive its distance through the expanding photosphere method
(EPM).  We present and discuss our photometric and spectroscopic observations
in \S~\ref{sec:photometry} and \S~\ref{sec:spectra}, respectively.  We estimate
the reddening of SN~1999gi from a variety of techniques in
\S~\ref{sec:reddening}.  We apply the EPM to SN~1999gi in
\S~\ref{sec:sn1999giepmdistance}, and compare the derived distance to existing
estimates of the distance to NGC~3184 in \S~\ref{sec:epmdistance}.  In
\S~\ref{sec:progenitormass} we discuss the impact of our results on the
progenitor mass limits previously determined by S01 and S02.  We summarize our
main conclusions in \S~\ref{sec:conclusions}.  Note that much of the background
material for the data and analysis presented in this paper, including the
details of the photometric and spectral reductions as well as many specifics of
the EPM technique itself, is thoroughly covered by earlier studies and
therefore not repeated here.  In particular, the recent analysis by L02 of
SN~1999em, a strikingly similar event to SN~1999gi, is frequently referenced.

\section{Reductions and Analysis}
\label{sec:obsandred}

\subsection{Photometry}
\label{sec:photometry}

We obtained 30 epochs of \bvri\ photometry for SN 1999gi, 26 of them taken with
the Katzman Automatic Imaging Telescope (KAIT; Filippenko et al. 2001) and 4
with the 1.2 m telescope at the Fred Lawrence Whipple Observatory (FLWO).  KAIT
is equipped with a $512\times 512$ pixel Apogee CCD camera (AP7) located at the
$f/8.17$ Cassegrain focus, providing a field of view of $6\farcm7 \times
6\farcm7$ with $0\farcs8$ per pixel.  The FLWO 1.2 m is a Ritchey-Chr\'{e}tien
reflector equipped with a ``4Shooter'' CCD mosaic camera (A. H. Szentgyorgyi et
al., in preparation), consisting of a $2 \times 2$ array of thinned, backside
illuminated, antireflection-coated Loral $2048 \times 2048$ pixel CCD detectors
situated at the $f/8$ Cassegrain focus, which provides a field of view
$11\farcm4 \times 11\farcm4$ on each chip.  All exposures with the mosaic were
made on the same CCD (Chip 3), and our observations were taken in a $2 \times
2$ binned mode, so that the resulting images were sampled at $0\farcs67$ per
pixel.  The seeing, estimated by the full width at half maximum (FWHM) of stars
on the CCD frame, generally ranged from $2\farcs5$ to $4\farcs5$ at KAIT, and
between $1\farcs5$ to $2\farcs5$ at FLWO.  Exposure times of 3 to 5 minutes
were typical at KAIT and 1 to 3 minutes at FLWO, with the longest exposures
taken in $B$.

\begin{figure}
\vskip -3.5in
\hskip -1.0in
\rotatebox{0}{
\scalebox{1.3}{
\plotone{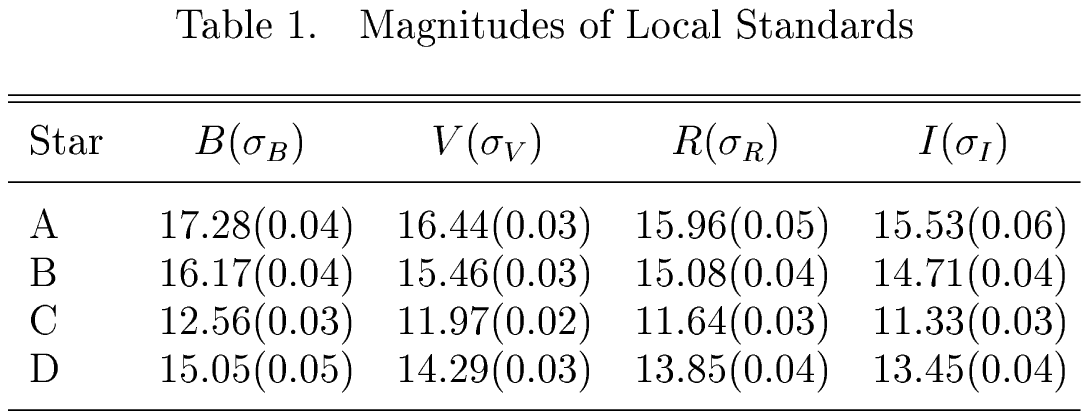}
} }
\end{figure}

\begin{figure}
\vskip -1.5in
\hskip -0.7in
\rotatebox{0}{
\scalebox{1.2}{
\plotone{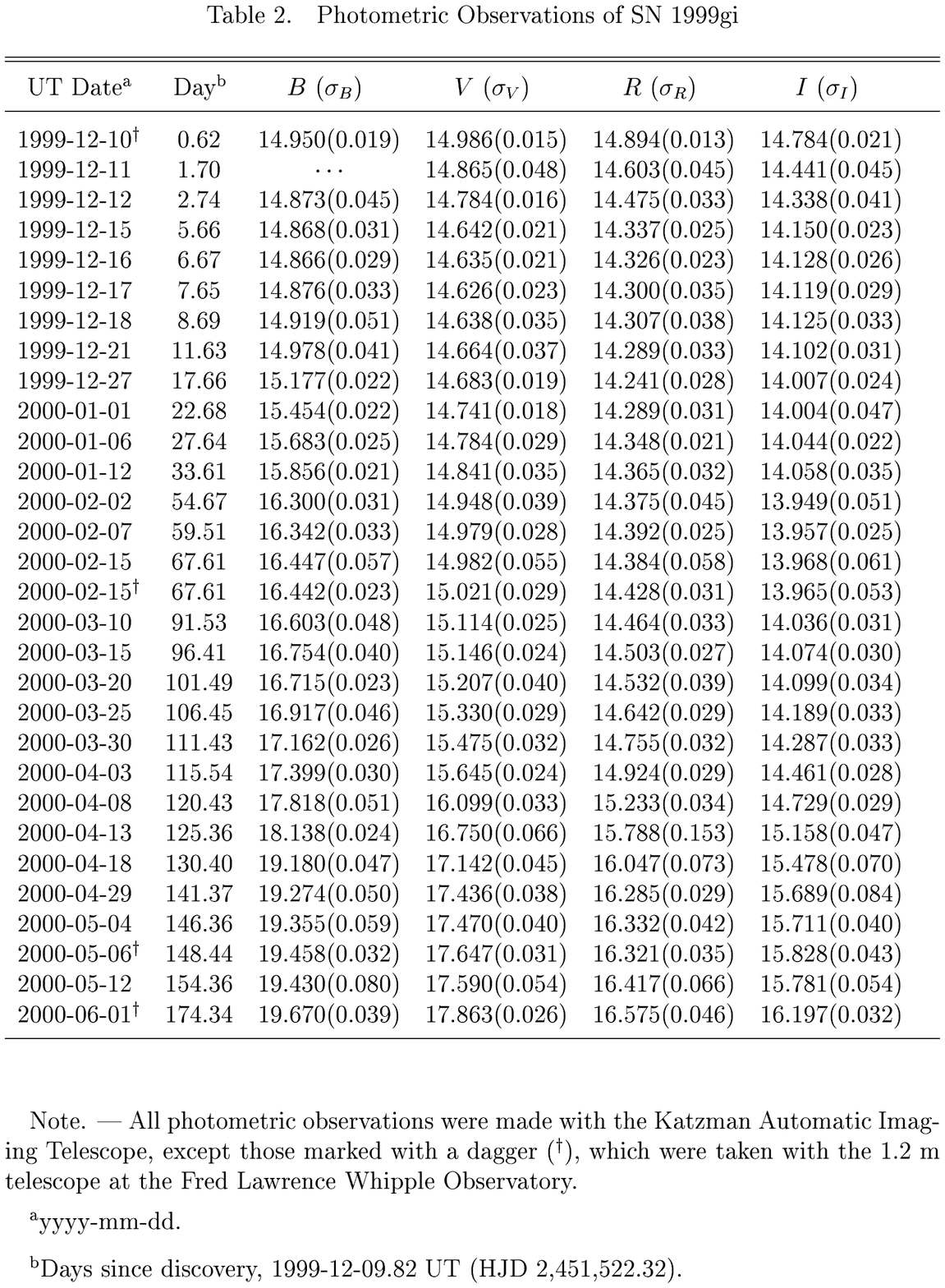}
} }
\end{figure}

\begin{figure}
\ssp
\begin{center}
\rotatebox{0}{
\scalebox{0.7}{
\plotone{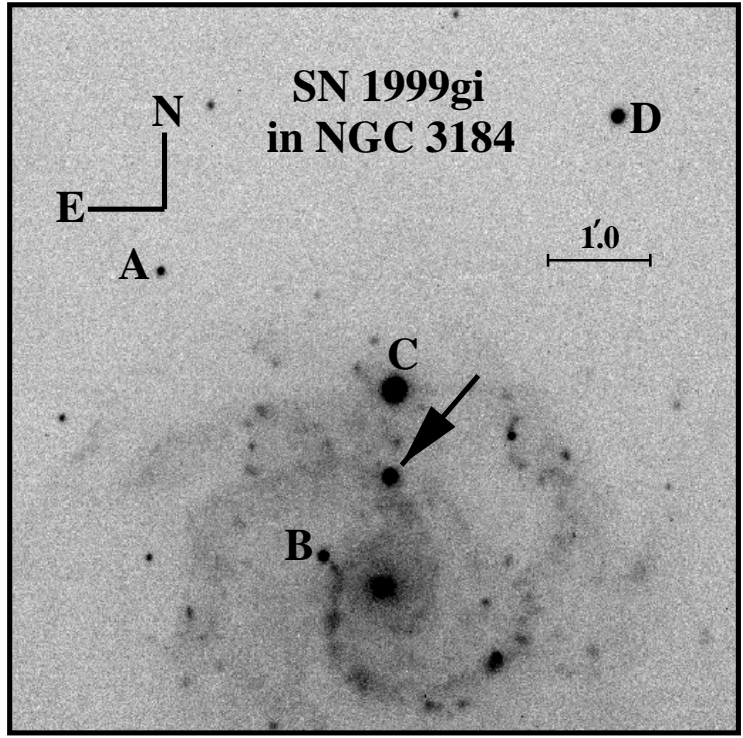}
}
}
\end{center}
\caption{$B$-band image of NGC 3184 taken on 1999 December 17 with the Katzman
Automatic Imaging Telescope (Filippenko et al. 2001),
with the local standards listed in Table 1 marked.  SN~1999gi ({\it arrow}) is
measured to be $60\farcs9$ north and $4\farcs7$ west of the center of NGC~3184. 
\label{fig:1} }
\end{figure}

Figure~\ref{fig:1} shows a KAIT $B$-band image of NGC 3184 taken on 1999
December 17.  The 4 ``local standards'' identified in the field of SN~1999gi
were used to measure the relative SN brightness on non-photometric nights.  We
reduced the data according to the procedure detailed by L02, except that
instead of using point-spread function (PSF) fitting to measure the SN's
brightness we employed the more accurate technique of galaxy subtraction (e.g.,
Filippenko et al. 1986; Richmond et al. 1995).  To use this technique, an image
of the host galaxy without the SN (i.e., a ``template'' image of the galaxy,
taken either before the explosion or after the SN has faded beyond detection)
is subtracted from the images containing the SN, with special care taken to
match the alignment, intensity, and PSFs of the images with and without the SN
present.  Since we did not possess deep filtered pre-explosion images of
NGC~3184, we obtained the templates on 17 January 2002 with the Nickel 1 m
reflector at Lick Observatory.  The Nickel telescope utilizes a Loral $2048
\times 2048$ pixel CCD, which we binned $2 \times 2$ to yield a plate scale of
$0\farcs37$ per pixel.

The absolute calibration of the field was accomplished on the three photometric
nights of 2000 April 3 (with KAIT) and 2002 January 17 and 18
(with the Nickel 1 m), by observing several fields of Landolt (1992)
standards over a range of airmasses in addition to the SN~1999gi field.  The
calibration of two of the local standards was further verified against the
photometric calibrations by S. Benetti (personal communication; star ``A'' in
Fig.~\ref{fig:1}) and B.
Skiff\footnote{\url{http://www.kusastro.kyoto-u.ac.jp/vsnet/Mail/vsnet-chat/msg02439.html}}
(star ``C'' in Fig.~\ref{fig:1}), and found to agree to within the reported
uncertainties.  We list the measured \bvri\ magnitudes, and the $1\ \sigma$
uncertainty (quadrature sum of the photometric error and the $1\ \sigma$
scatter of the individual photometric measurements taken on the three nights),
of the local standard stars in Table~1.

The transformation coefficients for the KAIT data to the standard
Johnson-Cousins (Johnson et al. 1966 for $BV$; Cousins 1981 for $RI$) systems
are those of L02.  For the FLWO data with Chip 3 of the 4Shooter, we derived
color terms from five photometric nights (1999 December 9, 2000 January 3,
January 5, February 7, and February 16) of
\begin{eqnarray}
B & = & b + 0.040(B - V) + C_B,\nonumber\\
V & = & v - 0.045(B - V) + C_V,\\
R & = & r - 0.080(V - R) + C_R,\nonumber\\
I & = & i + 0.028(V - I) + C_I,\nonumber
\label{eqn:photometric_solutions}\nonumber
\end{eqnarray}
\noindent where $bvri$ are the instrumental and \bvri\ the standard
Johnson-Cousins magnitudes. The terms $C_B, C_V, C_R,$ and $C_I$ are the
differences between the zero-points of the instrumental and standard
magnitudes, determined for each observation by measuring the offset between the
instrumental and standard magnitudes and colors of the local standard stars.

\begin{figure}
\ssp
\begin{center}
\rotatebox{0}{
\scalebox{1.0}{
\plotone{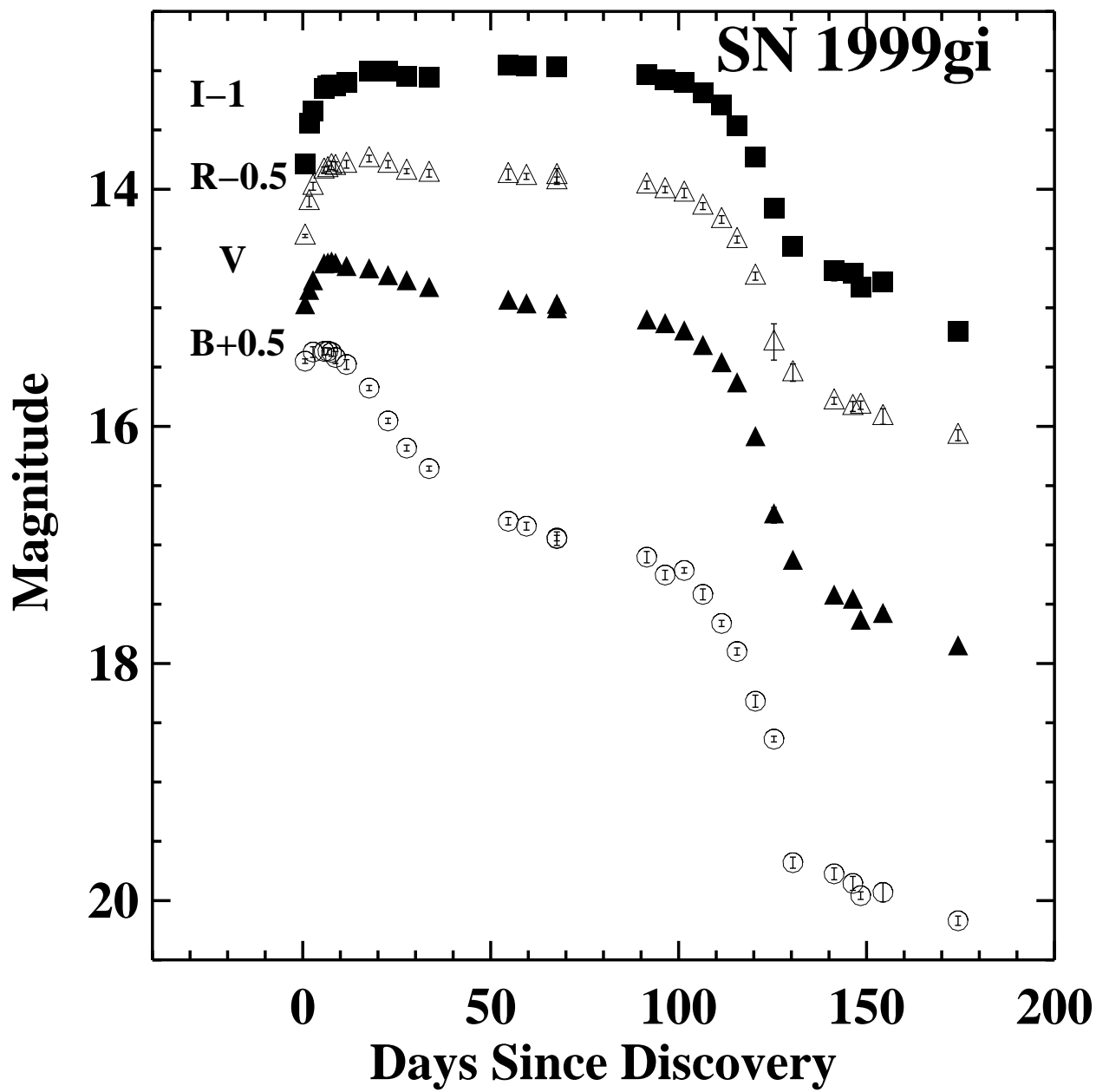}
}
}
\end{center}
\caption{\bvri\ light curves for SN 1999gi from Table~2.  For clarity, the
magnitude scales for {\it BRI} have been shifted by the amounts indicated.  In
most cases the error bar is smaller than the plotted symbol. 
\label{fig:2} }
\end{figure}

To transform the photometry to the standard Johnson-Cousins system we took the
weighted mean of the values individually derived from the four calibrator
stars.  Since star ``C'' was saturated in nearly all of the FLWO observations,
the FLWO photometry was usually derived using only three stars.  The results
of our photometric observations are given in Table 2 and shown in Figure
\ref{fig:2}.  The reported uncertainties come from the photometric and
transformation errors added in quadrature; in most cases, the total error was
dominated by the uncertainty in the transformation.


\begin{figure}
\ssp
\begin{center}
\rotatebox{0}{
\scalebox{1.0}{
\plotone{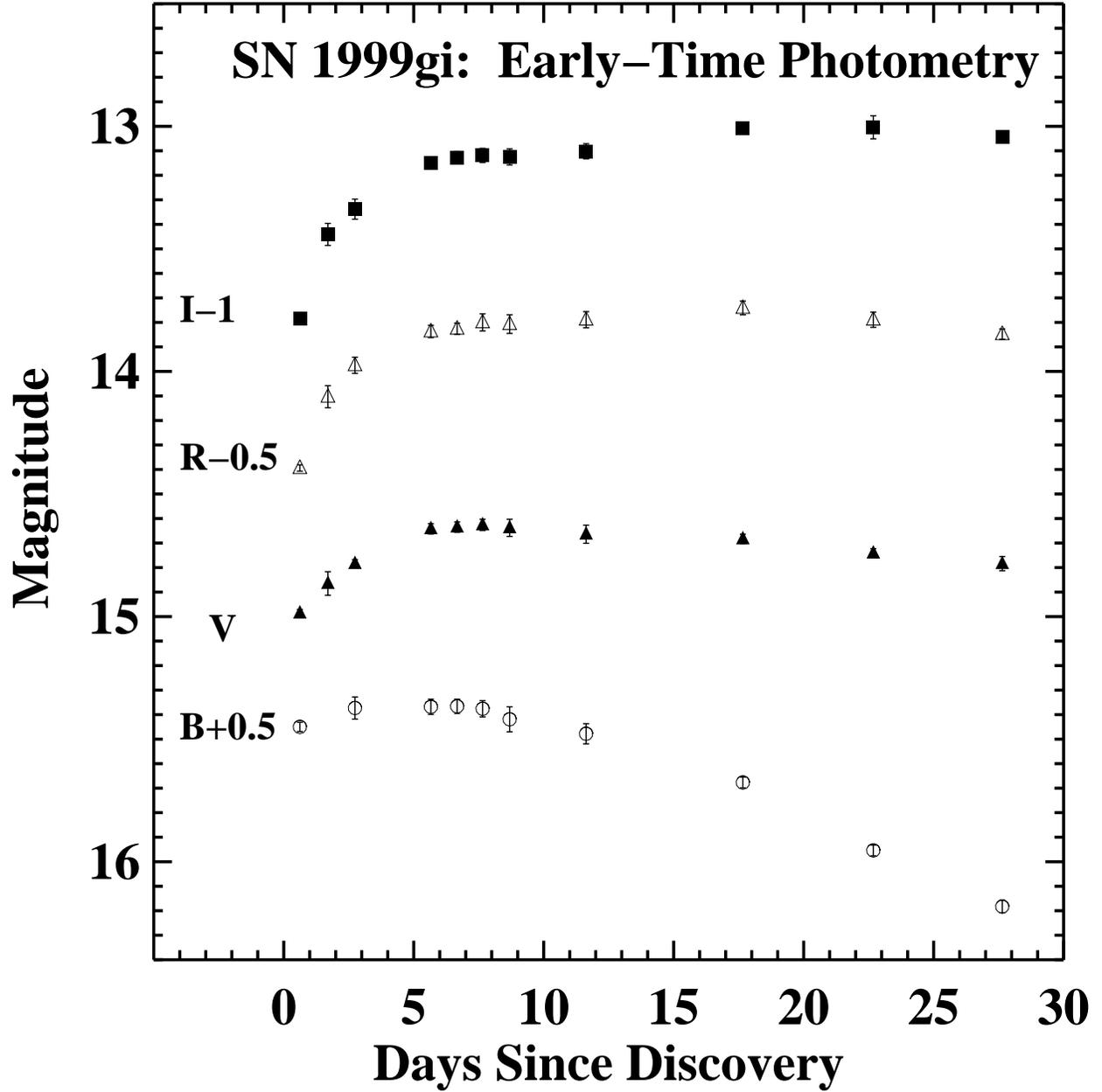}
}
}
\end{center}
\caption{A detail of the \bvri\ light curves presented in Figure~\ref{fig:2}
showing the first 30 days of photometric development for SN 1999gi. 
\label{fig:3} }
\end{figure}


\begin{figure}
\ssp
\begin{center}
\rotatebox{0}{
\scalebox{0.7}{
\plotone{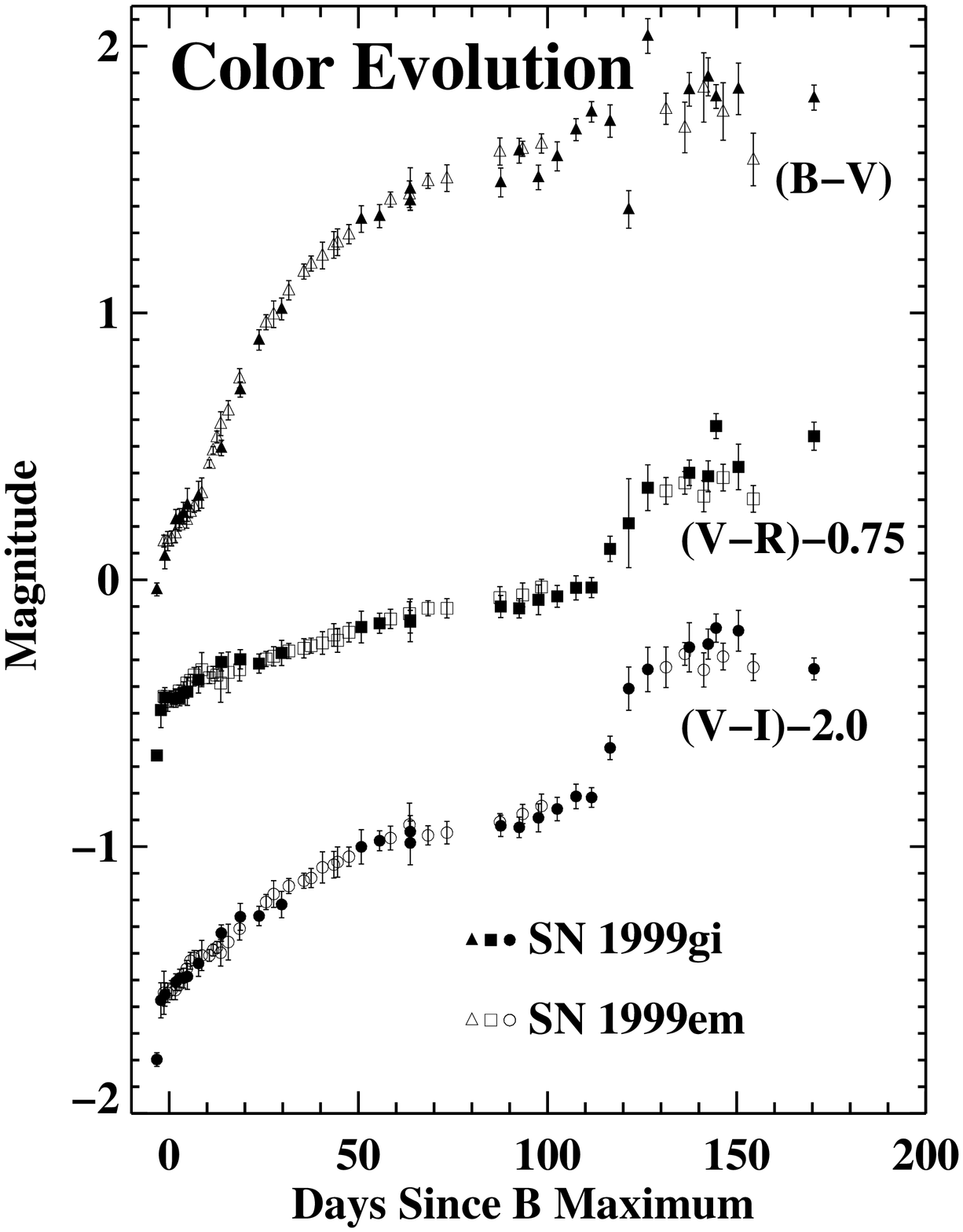}
}
}
\end{center}
\caption{\bv, \vr, and \vi\ color curves of SN~1999gi ({\it filled symbols})
compared with SN~1999em ({\it open symbols}; from L02), plotted against the
time since $B$-band maximum light ($3.9$ and $2.4$ days after discovery for
SN~1999gi and SN~1999em, respectively).  The data for SN~1999em have been
artificially reddened by an additional \ebv\ = 0.15 mag.  
\label{fig:4} }
\end{figure}

The light curves of SN~1999gi indicate a fairly rapid initial rise followed by
a clear plateau of nearly constant brightness in \vri\ that lasts until about
100 days after discovery.  To estimate the date of maximum in the $B$ and $V$
bandpasses (maximum brightness in $R$ and $I$ is not sharply defined and
actually comes significantly after the initial rise; see Fig.~\ref{fig:3} for
detail of the light curves near maximum light), we fit spline functions to the
data around the peak.  To estimate the uncertainty in the date of maximum we
measured the $1\ \sigma$ spread in the derived dates of maximum for 1000 sets
of synthetic photometric data, in which each epoch's photometry near the peak
was replaced with a value randomly chosen from a Gaussian-weighted distribution
centered on the values given in Table~2 and characterized by a standard
deviation equal to the uncertainty listed in Table~2.  From this analysis we
conclude that $m_B{\rm (max)} \approx 14.8$ mag occurred on HJD 2,451,526.2
$\pm\ 1.8$ (1999 December $13.7\pm 1.8$), which is $3.9 \pm 1.8$ days after
discovery.  For $V$ we find $m_V {\rm (max)} \approx 14.6$ mag on HJD
2,451,530.0 $\pm 1.6$ (1999 December $17.5 \pm 1.6$), which is $7.7 \pm 1.6$
days after discovery.  The decline in the $B$ band over the first 100 days
after maximum light is $\beta^B_{100} = 1.9 $ mag, which definitively
establishes SN~1999gi as a Type II-P event according to the definition by Patat
et al. (1994), used to discriminate SNe~II-P ($\beta^B_{100} < 3.5$
mag) from SNe~II-L ($\beta^B_{100} > 3.5$ mag).  The average apparent $V$-band
brightness during the plateau, defined as the unweighted mean of the values
from days $20 {\rm\ to\ } 100$ after the explosion (determined by the EPM
analysis in \S~\ref{sec:sn1999giepmdistance}), is $\overline{m}_V~{\rm
(plateau)} = 14.90$ mag.

The photometric behavior of SN~1999gi is extremely similar to that of
SN~1999em, with both having plateau durations of $\sim 95$ days after $B$-band
maximum as well as similar $B$-band decline rates (see L02).  In fact, once an
allowance is made for a small reddening difference ($\Delta\EBV \equiv
(\EBV_{\rm 99gi} - \EBV_{\rm 99em}) = 0.15$ mag), the color evolution of the
two objects is nearly identical (Fig.~\ref{fig:4}).  The optical ``double
peak'' noted by L02 for SN~1999em, in which a second local maximum follows the
initial peak ($8.6 \pm 2.3$ days later for the $V$-band in SN~1999em), is also
discernable in the photometry of SN~1999gi (see Fig.~\ref{fig:3}). However, the
second peak is certainly not as obvious as it was in the photometry of
SN~1999em, a situation likely exacerbated by the relatively sparse photometric
coverage obtained during this period for SN~1999gi compared with SN~1999em.

\subsection{Spectroscopy}
\label{sec:spectra}

Table 3 lists the spectral observations of SN~1999gi.  All one-dimensional
sky-subtracted spectra were extracted optimally (Horne 1986) in the usual
manner.  Each spectrum was then wavelength and flux calibrated, as well as
corrected for continuum atmospheric extinction and telluric absorption bands
(Wade \& Horne 1988; Bessell 1999; Matheson et al. 2000).  With the exception
of the very first spectrum (day 0.61, 300 s observation), all spectra were
observed near the parallactic angle (Filippenko 1982), so the spectral shape
should be quite accurate.\footnote{A direct comparison of the first spectrum
with the one taken immediately after it at the parallactic angle (day 0.62, 600
s observation) confirms that the low airmass (1.14) of this observation renders
the effects of differential light loss negligible.}

\begin{figure}
\vskip -1.in
\hskip -2.in
\rotatebox{180}{
\scalebox{1.1}{
\plotone{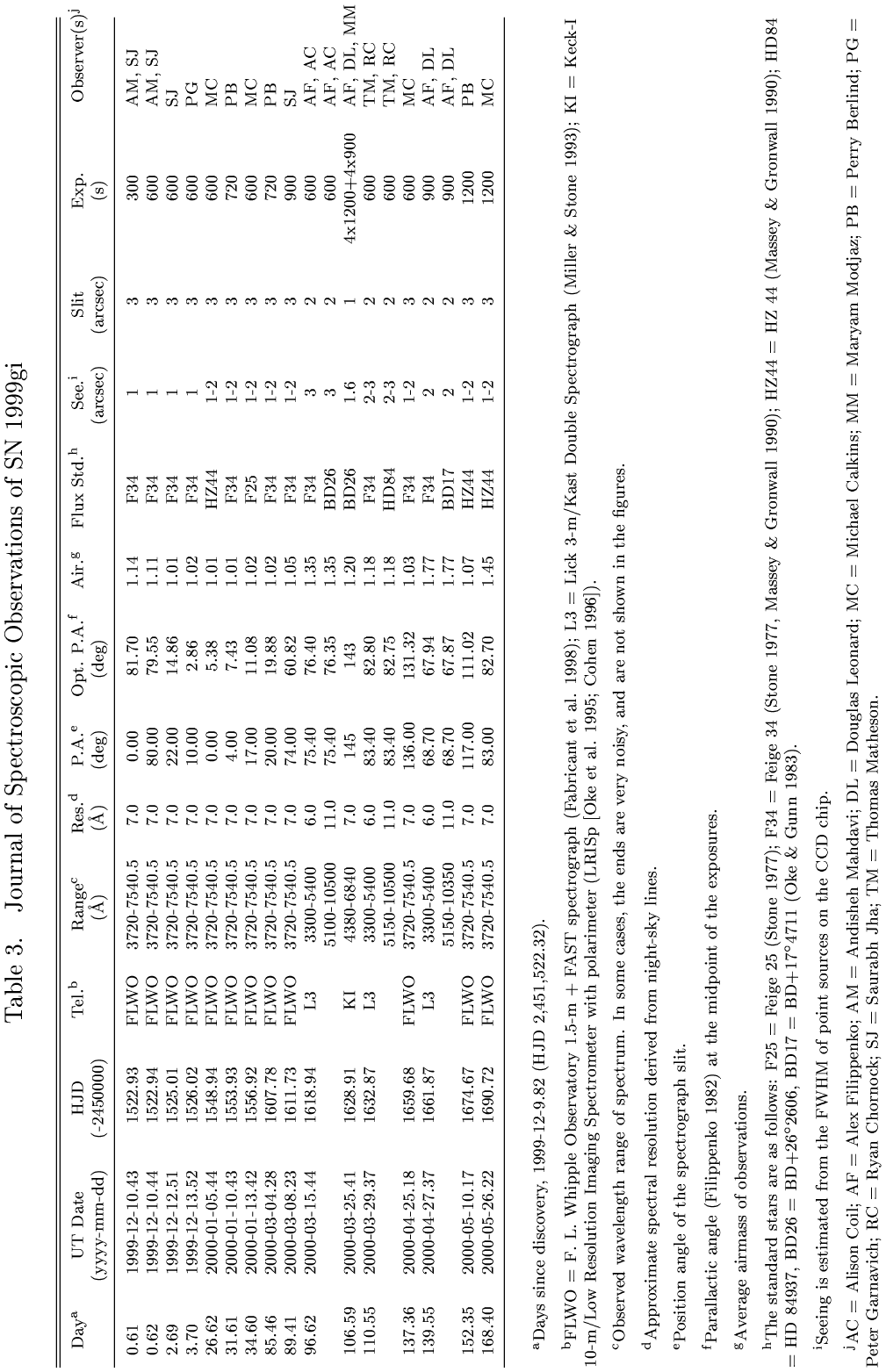}
} }
\end{figure}

\begin{figure}
\ssp
\begin{center}
\rotatebox{0}{
\scalebox{1.0}{
\plotone{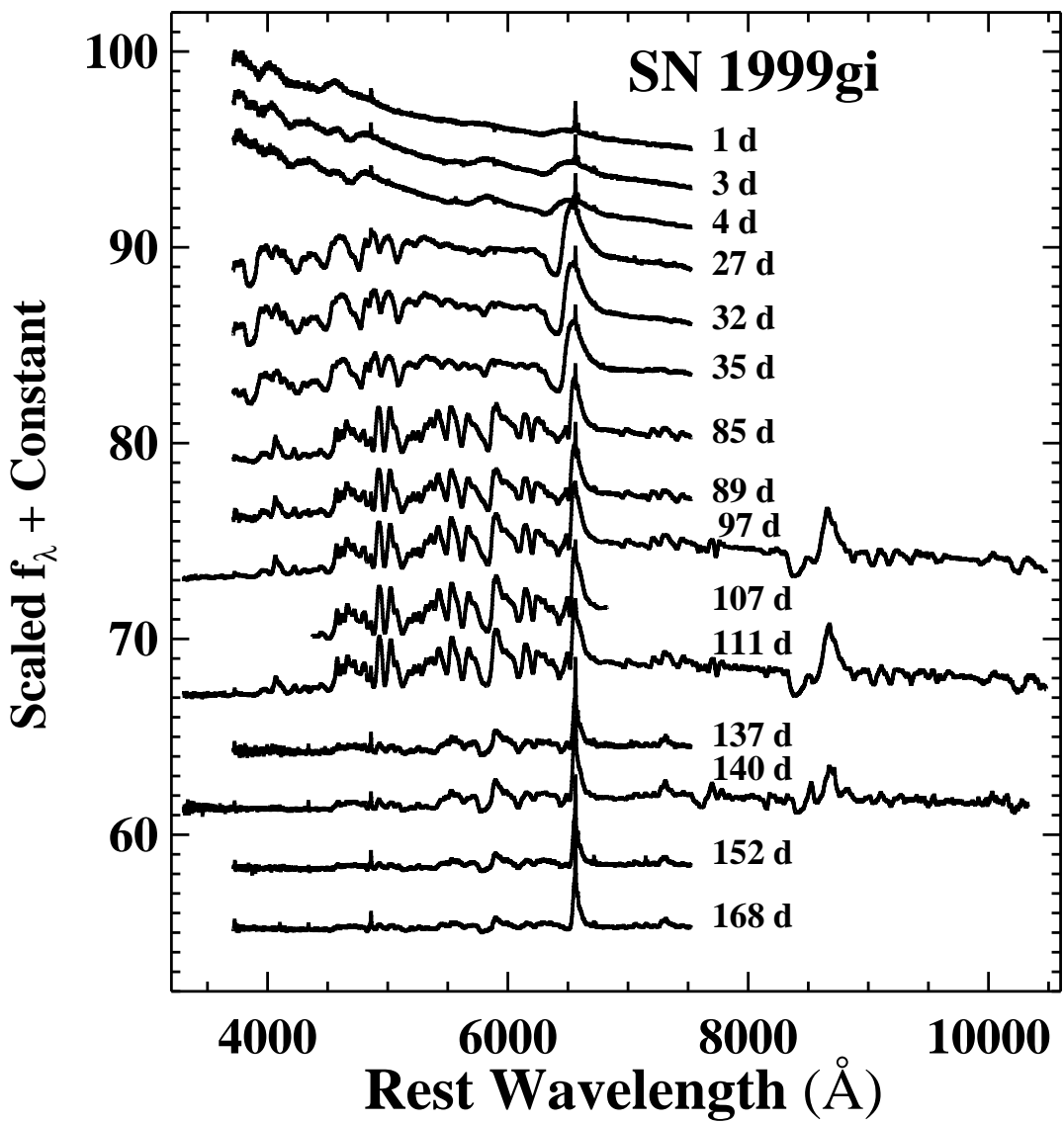}
}
}
\end{center}
\caption{Optical flux spectra of SN~1999gi with day since discovery indicated.
Note the narrow emission spike evident near the peaks of the H$\alpha$
($6563$ \AA) and \hbeta\ ($4861$ \AA) profiles due to a
superposed \ion{H}{2} region.  In this and all figures a recession velocity of
552 \kms\ has been removed from the observed spectrum (\S~\ref{sec:spectra}).
\label{fig:5} }
\end{figure}

The spectral evolution of SN~1999gi during the first 168 days of its
development is shown in Figure~\ref{fig:5}, and is very similar to that seen
for SN~1999em (L02; see also Leonard \& Filippenko 2001, Fig.~10).  As is
typical for SNe II-P, the early-time spectrum is characterized by a smooth
thermal continuum with superposed hydrogen Balmer and \ion{He}{1} $\lambda
5876$ P-Cygni lines.  During the plateau phase, numerous metal-line P-Cygni
features complicate the spectrum (see L02 for a complete identification of
these features).  Emission-dominated lines then become prominent as SN~1999gi
drops off the plateau and makes the transition to the nebular phase.


To estimate the recession velocity of SN~1999gi we measured the peak wavelength
of the narrow \halpha\ line resulting from the superposed \ion{H}{2} region.
From this, we determine that $v_{\rm rec} = 552 \pm 10$ \kms, where the
uncertainty is the $1\ \sigma$ spread in the velocities derived from the
individual spectra.  This velocity is to be compared with the value reported
for NGC~3184 by Strauss et al. (1992) of $592 \pm 1$ \kms, which was derived
from narrow \ion{H}{2} emission in the central $10\arcsec$ of the galaxy.
Since SN~1999gi is quite far away from the center of NGC~3184
(Fig.~\ref{fig:1}), our velocity is likely a more accurate estimate of the true
velocity of the SN.  We therefore use it to derive all photospheric velocities
in the EPM analysis in \S~\ref{sec:sn1999giepmdistance}; ultimately, using the
larger Strauss et al. (1992) value would result in a final EPM distance that is
$\sim 1\%$ larger than that derived with our lower recession velocity.

\begin{figure}
\ssp
\begin{center}
\rotatebox{0}{
\scalebox{0.7}{
\plotone{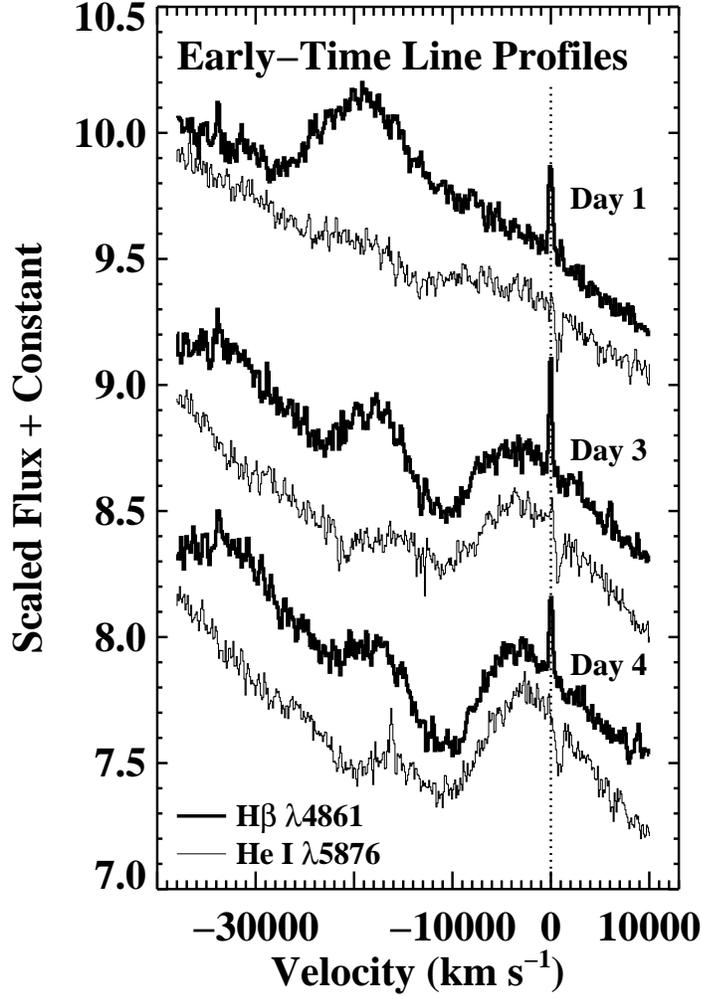}
}
}
\end{center}
\caption{The early-time development of the H$\beta$ ({\it thick lines}) and
\ion{He}{1} $\lambda5876$ ({\it thin lines}) profiles, with days since
discovery indicated.  Note the high-velocity ($v > 20,000$ \kms) P-Cygni
absorption that exists in addition to the ``normal,'' lower-velocity ($v
\approx 10,000$ \kms) profiles in both lines.  In the first spectrum, in fact,
the high-velocity component of \hbeta\ dominates over the weaker, low-velocity
profile.  Note also the narrow emission spike at zero velocity in the H$\beta$
profiles due to a superposed \ion{H}{2} region, the sharp, narrow absorption
feature just redward of zero velocity in the \ion{He}{1} $\lambda5876$ spectra
from \ion{Na}{1} D gas in NGC~3184, and the small spike at $v \approx -16,000$
\kms\ in the day 4 \ion{He}{1} $\lambda 5876$ profile from poor telluric
[\ion{O}{1}] $\lambda 5577$ removal.
\label{fig:6}  }
\end{figure}

The very early-time \hbeta\ and \ion{He}{1} $\lambda 5876$ line profiles of
SN~1999gi are shown in Figure~\ref{fig:6}, and warrant special comment.  In
early-time spectra of SN~1999em, Baron et al. (2000) identify ``complicated
P-Cygni'' profiles, in which the usual P-Cygni absorptions for \hbeta\ and
\ion{He}{1} $\lambda 5876$ are accompanied by a second absorption at much
higher velocity.  From comparison with theoretical models, Baron et al. (2000)
conclude that these high-velocity absorption features result from non-local
thermodynamic equilibrium effects that produce two line-forming regions in the
expanding atmosphere.  Leonard et al. (2001) also identify these high-velocity
features in an early spectrum of SN~1999em.  In addition, L02 propose that
complex P-Cygni profiles of strong lines may also explain a number of
previously unidentified absorption features in spectra of SN~1999em at later
times, during the photospheric and early nebular phases; detailed modeling,
however, is required to confirm these identifications.  The early-time
absorption features observed in the spectra of SN~1999em were rather subtle,
and left open the question of whether they were true P-Cygni features
containing both a dip and a peak.  As clearly seen in Figure~\ref{fig:6},
similar high-velocity features are also evident in the early spectra of
SN~1999gi.  Furthermore, the \hbeta\ profiles from days 3 and 4 reveal that the
high-velocity feature is indeed characterized by a true P-Cygni profile.

The earliest spectrum, from day 1, is particularly interesting.  In \ion{He}{1}
$\lambda 5876$ there is just a hint of the higher-velocity component in
addition to the ``normal'' line feature.  Near the expected location of the
high-velocity feature in \hbeta\ (i.e., based on its position in the spectra
from days 3 and 4), there is a strong emission feature which, if identified
with \hbeta, extends out to nearly $-30,000$ \kms.  If this association is
confirmed by detailed models, it would demonstrate the existence of very high
velocity material in the outer envelope of SN~1999gi at these early times.
Although the ``normal,'' lower-velocity component of \hbeta\ is likely present
at this early epoch, it is clearly much weaker than the higher-velocity
component.\footnote{Although not shown in the figure, we note that there is
also a strong feature at the expected location of a high-velocity component for
H$\gamma$, but the identification is ambiguous since the lower-velocity
absorption of H$\delta$ is expected at nearly the same location.}
Interestingly, near the expected location of \halpha, we find that while the
``normal'' profile is observed at all epochs (including day 1), a high-velocity
feature is {\it not} apparent in the spectra from days 1, 3, and 4.  At later
times, during the photospheric and early nebular phases, the high-velocity
absorption features attributed by L02 to \hbeta, \ion{Na}{1} D, and \halpha\ in
spectra of SN~1999em are also seen in the spectra of SN~1999gi: They are
consistently detected in \hbeta\ and \ion{Na}{1} D at $9000 \lesssim v \lesssim
11,000$ \kms\ through the spectrum on day 111, and are obvious in \halpha\ (at
$v \approx 13,700 $ \kms) in the spectra on days 32 and 35.  Further detailed
modeling is urged in order to better establish the physical mechanism behind
these intriguing features.


\subsection{The Reddening of SN~1999gi}
\label{sec:reddening}

SN~1999gi has minimal Galactic reddening, \ebv\ = 0.017 mag (Schlegel,
Finkbeiner, \& Davis 1998).  In order to estimate the {\it total} reddening (or
to set limits on how high it can be), we shall compare the results obtained
from the following five independent techniques. \\

\begin{figure}
\ssp
\begin{center}
\rotatebox{0}{
\scalebox{0.65}{
\plotone{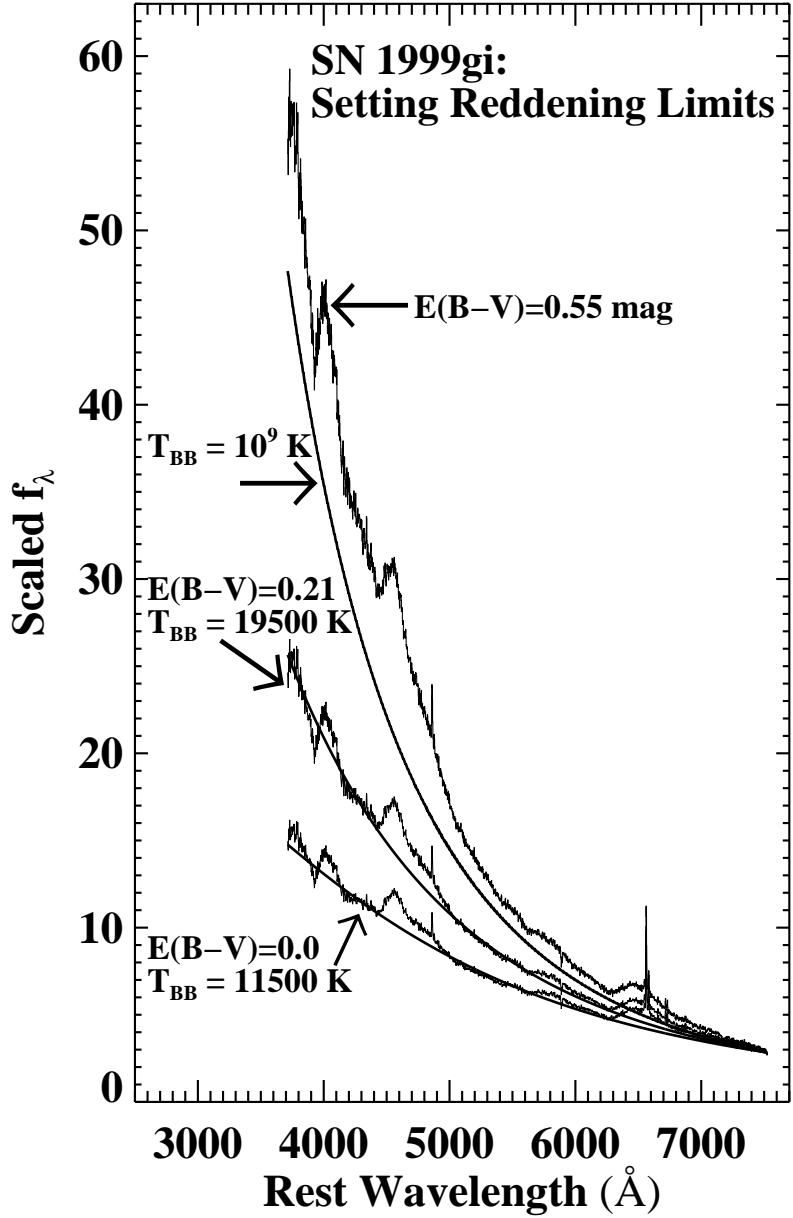}
}
}
\end{center}
\caption{Total-flux spectrum of SN~1999gi taken 0.62 days after discovery,
dereddened by \ebv = 0.55, 0.21, and 0.00 mag ({\it top, middle}, and {\it
bottom}, respectively), with blackbody functions at the indicated temperatures
also shown.  For assumed reddening values of $\Ebv < 0.45$ mag, it is possible
to trade increases in reddening with increases in temperature.  However, when
the spectrum is dereddened by $\Ebv > 0.45$ mag (represented here by $\EBV =
0.55$ mag), the spectral slope of SN~1999gi cannot be matched by a blackbody at
{\it any} temperature, no matter how hot (represented here by $10^9$ K).  The
inability of an arbitrarily hot blackbody to match the continuum slope for
$\Ebv > 0.45$ mag effectively establishes 0.45 mag as an upper reddening limit.
For ease of comparison, the lower two spectra have been scaled to match the
flux level of the top spectrum at the red end.
\label{fig:7}  }
\end{figure}

\indent 1. {\it Compare the early-time spectral shape with blackbody
functions.}  As discussed by Eastman et al. (1996), an upper reddening limit
can be established for an SN II-P by comparing its very early-time spectral
shape with an arbitrarily hot blackbody function.  The basic idea is that at
very high temperatures (i.e., early times for an SN II-P) the optical continuum
is on the Rayleigh-Jeans tail of the Planck spectrum, whose slope is quite
insensitive to temperature (i.e., in the limit $T \rightarrow \infty, \partial
\ln f_\lambda / \partial \ln \lambda \approx -4$).  If a hot, early-time
spectrum (with accurate continuum shape) is available (the earlier, the better,
since cooler spectra will result in a less-restrictive upper bound), one can
determine the maximum allowable reddening by dereddening the spectrum to the
point at which even an unphysically hot blackbody function (say, $10^9$ K) can
no longer fit the continuum.  As shown in Figure~\ref{fig:7}, the inability of
a blackbody at any temperature to match the continuum shape of SN~1999gi when
dereddened by $\Ebv > 0.45$ mag effectively establishes this as an upper limit
for the reddening of SN~1999gi.  The true reddening of the SN, of course, is
likely to be substantially lower.  For instance, the more reasonable color
temperature of $19,500$ K is shown in Figure~\ref{fig:7} to correspond with a
reddening of $\Ebv = 0.21$ mag.


\indent 2. {\it Compare the color evolution of SN~1999gi with SN~1999em.}
Schmidt, Kirshner, \& Eastman (1992) demonstrate that the color evolution of
the 5 SNe II-P in their study is quite similar, with the observed discrepancies
consistent with reddening differences among the objects.  Although still
limited by the small number of well-studied examples, it is tempting to posit
that all SNe II-P undergo similar color evolution, at least through the end of
the recombination phase.  If this is so, then it should be possible to estimate
(and set hard upper limits on) the reddening of SN~1999gi by comparing it with
SN~1999em, which was extensively modeled by Baron et al. (2000), who set a firm
upper limit of $\Ebv_{\rm tot} < 0.15$ mag and estimated the likely reddening
to be $0.05 \lesssim \Ebv_{\rm tot} \lesssim 0.10$ mag.  (The EPM analyses of
Hamuy et al. 2001 and L02 are also consistent with low reddening for
SN~1999em.)  The similar color evolution shown in Figure~\ref{fig:4}, where the
data for both objects are plotted with respect to the date of $B$-band maximum,
suggests a reddening difference between the two objects of $\Delta\Ebv = 0.15$
mag, implying that $0.20 \lesssim \Ebv_{\rm 99gi} \lesssim 0.25$ mag, with
$\Ebv_{\rm 99gi} < 0.30$ mag serving as a hard upper limit.  Of course, this
analysis assumes the intrinsic color evolution of the two objects to be
identical with respect to the date of maximum brightness.  If the analysis is
instead done with respect to the explosion date (derived from the EPM
analyses), then the best agreement is found for $0.15 \lesssim \Ebv_{\rm 99gi}
\lesssim 0.20$ mag, in which case $\Ebv < 0.25$ mag then becomes the upper
bound.  We therefore conclude that the comparison of the color evolution of
SN~1999gi with that of SN~1999em implies a reddening of $\Ebv \approx 0.20$ mag
for SN~1999gi.

\indent 3. {\it Translate observed \ion{Na}{1} D interstellar absorption
equivalent width into a reddening estimate.}  The correlations between the
\ion{Na}{1} D $\lambda\lambda 5890, 5896$ interstellar line equivalent width
and reddening found by Barbon et al. (1990) and Munari \& Zwitter (1997) are
known to have a very large scatter, especially when only low-resolution spectra
are available (see Leonard \& Filippenko 2001 for a more extensive discussion
of the utility of these relations).  Nevertheless, it is useful to compare the
predictions that these relations make with the results of the other, perhaps
more accurate, techniques.

To derive the \ion{Na}{1} D equivalent width in the spectrum of SN~1999gi, we
first co-added the normalized region around the \ion{Na}{1} D lines in the 5
earliest spectra, which had the highest signal-to-noise ratio (S/N) in the
region of interest (i.e., we combined the spectra up through and including the
one taken on day 31.61; the spectrum from day 34.60 was corrupted by a
cosmic-ray spike in the \ion{Na}{1} D region, and the high S/N spectrum from
day 106.59 could not be used with confidence since the \ion{Na}{1} D absorption
occurred very near the maximum of an SN feature, making proper continuum
placement difficult).  From this composite spectrum we measure $W_\lambda {\rm
(Na\ I\ D)} = 0.83 \pm 0.15$~\AA, with $W_\lambda {\rm (Na\ I\
D1[\lambda5896])} = 0.31 \pm 0.15$~\AA, and $W_\lambda {\rm (Na\ I\
D2[\lambda5890])} = 0.47 \pm 0.20$~\AA; the uncertainty in the measurements
represents the $1\ \sigma$ spread in the automated measurements of 1000 sets of
simulated data characterized by the S/N of the composite spectrum.  To solve
for the D1 and D2 components separately we used the Levenberg-Marquardt
maximization algorithm (Press et al. 1992) with Gaussians constrained to have
identical widths (set equal to the spectral resolution), and separated by
$5.97$ \AA.  Note that since the $W_\lambda {\rm (Na\ I\ D1[\lambda5896])}$ and
$W_\lambda {\rm (Na\ I\ D2[\lambda5890])}$ values are the result of model fits,
whereas $W_\lambda {\rm (Na\ I\ D)}$ is measured directly from the (noisy)
spectrum, the sum of the equivalent widths of the individual lines does not
exactly equal the reported total equivalent width (but is equal within the
uncertainty).  From these values, the Barbon (1990) relation yields $\Ebv_{\rm
host} \approx 0.21$ mag, and the Munari \& Zwitter (1997) relation gives
$\Ebv_{\rm host} \approx0.27$ mag.  From a similarly constructed composite
spectrum centered on the expected location of the \ion{Na}{1} D lines in the
Milky Way (i.e., at zero velocity), the absence of detectable absorption
implies an upper limit on the total equivalent width due to Galactic gas of
$W_\lambda(3\sigma) = 0.06$ \AA\ through Equation~4 of Leonard \& Filippenko
(2001); this translates into an upper reddening limit of $\Ebv_{\rm MW} < 0.02$
mag from the Barbon (1990) relation, which concurs with the value given by
Schlegel et al. (1998).  The total reddening predicted by these crude relations
is therefore found to be $\Ebv \approx 0.25$ mag, a value that is consistent
with the upper limits and values derived by the other methods.

\indent 4.  {\it Assume the reddening to SN~1999gi to be the same as that
derived to the young OB association in which it is situated.}  From analysis of
WFPC2 {\it HST} observations of NGC~3184, S01 conclude that SN~1999gi originates in a
resolved, young OB association.  By comparing the evolutionary isochrones of
Lejuene \& Schaerer (2001) with the color-magnitude diagram of the stars in the
OB association (i.e., those within $12\farcs5$ of SN~1999gi), S01 find $\Ebv =
0.15$ mag to give the best fit to the cluster, and note that for reddenings above
$\Ebv = 0.30$ mag the isochrones fail to match the color of the brightest sources.  

\indent 5.  {\it Assume the reddening to SN~1999gi to be the same as that
derived to a nearby \ion{H}{2} region.}  In order to estimate nebular
abundances in NGC~3184, Zaritsky, Kennicutt, \& Huchra (1994) determined the
reddening towards 19 \ion{H}{2} regions, one of which is just $8\farcs5$ away
from SN~1999gi, and located in the OB association identified by S01 to likely
contain the progenitor of SN~1999gi. Assuming $R_V = 3.1$ (Savage \& Mathis
1979), Zaritsky et al. (1994) find $\Ebv = 0.34$ mag to this nearby \ion{H}{2}
region.

None of the techniques that we have used to estimate the total reddening to
SN~1999gi is individually unassailable.  Potential weaknesses of each technique
are as follows (given for each of the techniques, respectively). (1) Early-time
SN II-P color may depart slightly from a single-temperature blackbody. (2) All
SNe II-P may not evolve in a similar manner, and SN~1999gi and SN~1999em may
not have been identical events. (3) Sodium is known to be only a fair tracer of
dust, and our low-resolution spectra do not resolve the individual absorption
systems that contribute to the \ion{Na}{1} D lines, making it impossible to
estimate the effect of line saturation. (4 \& 5) The reddening along the
specific line-of-sight to SN~1999gi may differ from the average reddening to
the OB association and/or the reddening to specific association members (see,
e.g., Yadav \& Sagar 2001).  Nonetheless, taken together the general agreement
of all the estimates makes a compelling case for low reddening to SN~1999gi,
with $\Ebv = 0.20$ mag (comparison with SN~1999em); $0.25$ mag (\ion{Na}{1} D
absorption); $0.15$ mag (reddening to OB association); and $0.34$ (reddening to
\ion{H}{2} region), and an upper reddening limit of $\Ebv < 0.45$ mag coming
from the analysis of the early-time continuum shape.  If the color evolution of
SN~1999gi is intrinsically the same as that of SN~1999em with respect to the
time of maximum light, then we can restrict the upper bound even further, to
$\Ebv < 0.30$ mag.  In \S~\ref{sec:sn1999giepmdistance}, we shall see that
additional evidence for low reddening to SN~1999gi comes from the EPM analysis
itself, which predicts $\Ebv \approx 0.10$ mag.  Taking the simple average of
the five specific reddening estimates (i.e., including the EPM estimate derived
in \S~\ref{sec:sn1999giepmdistance}, but not incorporating the upper limit
derived from the blackbody comparison) yields $\Ebv = 0.21 \pm 0.09$ mag, where
the uncertainty is the $1\ \sigma$ spread of the individual reddening estimates
from the mean.  We adopt this value as our best estimate of the reddening of
SN~1999gi.

\section{The EPM Applied to SN~1999gi}
\label{sec:sn1999giepmdistance}

To derive the EPM distance to SN~1999gi we follow the procedure detailed by
L02.  Briefly, an EPM distance is calculated by comparing the linear radius of the
expanding supernova photosphere, $R$, with the photosphere's angular size,
$\theta$, to derive the distance to the SN, $D$ (Kirshner \& Kwan 1974).  The radial velocity of the
expanding photosphere, $v$, is found from the Doppler shifting of the spectral
lines, so that $R = v(t - t_{\circ})$, where ($t - t_{\circ}$) is the time
since explosion and the SN is assumed to be in free expansion.  The
photosphere's theoretical angular size, $\theta$, is calculated by comparing
the observed flux with that predicted from theoretical models (i.e., a
``dilute'' blackbody; Eastman et al. 1996) for a spherical SN photosphere as

\begin{equation}
\theta = \sqrt{\frac{f_\nu 10^{0.4A_\nu}}{\zeta_\nu^2 (T_c) \pi B_\nu (T_c)}} ,
\label{eq:5}
\end{equation}

\noindent where $B$ is the Planck function at color temperature $T_c$, $f_\nu$
is the flux density received at Earth, $A$ is the extinction, and $\zeta(T_c)$
is the color temperature dependent ``dilution factor'' (or ``distance
correction factor,'' since it ``corrects'' derived distances such that $D_{\rm
actual} = \zeta D_{\rm measured}$; see L02 for a thorough discussion of this
term).  Since accurate spectrophotometry is generally not available,
Equation~\ref{eq:5} is typically recast in terms of broadband photometry, with
$T_c$ and $\zeta$ derived for some subset of $BVIJHK$.  With $R$ and $\theta$
known, $D$ can be found since $\theta = R/D$ in the small-angle approximation.
Substituting $v(t - t_{\circ})$ for $R$ and rearranging, we arrive at
\begin{equation}
t = D\left(\frac{\theta}{v_{\rm phot}}\right) + t_\circ .
\label{eq:2}
\end{equation}

\noindent A plot of $\theta/v_{\rm phot}$\ against $t$ should therefore result
in a line with slope $D$ and $y$-intercept $t_{\circ}$.

As discussed by L02, the overall level of the theoretically derived dilution
factor, $\zeta (T_c)$, has long been touted as potentially the largest source
of {\it systematic} uncertainty in the application of the EPM to SNe II-P, due
mainly to the fact that most applications of the EPM have relied on the
dilution factors produced by only one modeling group (i.e., Eastman et
al. 1996).  Since no other independent modeling group has published dilution
factors appropriate for SNe II-P,\footnote{Eastman et al. (1996) stress that
their dilution factors are {\it only} appropriate for SNe II-P, and not to
peculiar variants of the SN II subclass.  Eastman et al. (1996) also note that
the models of Montes \& Wagoner (1995), while appropriate for SNe II-P, are
difficult to directly compare with their results, since Montes \& Wagoner
(1995) define $\zeta$ somewhat differently, and also do not model the effect
that line blanketing has on the apparent color temperature.}  it is difficult
to quantify the amount of systematic uncertainty that the reliance on only one
group's dilution factors contributes to EPM distances.  However, to get a sense
of the possible differences among modeling groups, it is instructive to
consider the two {\it unusual} (i.e., not SNe II-P) core-collapse events for
which independent dilution factors from another group have been derived:
SN~1987A and SN~1993J.

Recently, Mitchell et al. (2002) derive a distance to SN~1987A of $D = 50 \pm
5$ kpc through detailed modeling of the spectra and the use of the
``spectral-fitting expanding atmosphere method'' (SEAM; Baron et al. 1993;
Baron, Hauschildt, \& Branch 1994; Baron et al. 1995).  This distance is in
excellent agreement with other, independent distance estimates to SN~1987A and
the Large Magellanic Cloud (see, e.g., Gibson 2000).  SN~1987A was not a
classical SN II-P: Its progenitor was a compact blue supergiant and its light
curve became powered by radioactivity substantially earlier than is typical for
SNe II-P, which result from progenitors with more extended envelopes.
Nonetheless, Mitchell et al. (2002) do provide a direct comparison between the
dilution factors resulting from their SN~1987A-like models with two of the SN
II-P models by Eastman et al. (1996), for dilution factors derived using the
$BV$ bandpasses to estimate the color temperature.  While they find quite good
agreement between the dilution factors at both high ($T \gtrsim 8000$ K) and
low ($T \lesssim 5000$ K) temperatures, they note some disagreement in the
important temperature range of $5000 - 8000$ K. In this region, the dilution
factors of Mitchell et al. (2002) are $\sim 50 - 60\%$ larger than those of
Eastman et al. (1996). This implies that using the dilution factors of Eastman
et al. (1996), then, would result in a shorter distance to SN~1987A (e.g.,
$50\% - 60\%$ smaller if only this temperature range were considered) than that
found by using the Mitchell et al. (2002) models.  Recently, Hamuy (2001) has
derived the EPM distance to SN~1987A using the dilution factors of Eastman et
al. (1996), as slightly modified for the effects of telluric absorption by
Hamuy et al. (2001).  As anticipated from the comparison by Mitchell et
al. (2002), Hamuy (2001) indeed finds a shorter distance to SN~1987A.  Using
the dilution factors for the $BV$ filter combination (while noting that these
dilution factors are not entirely appropriate for SN~1987A), Hamuy (2001) finds
$D = 37$ kpc.  The discrepancy between the two results may be due to the
unusual nature of SN~1987A, and not to fundamental differences in the modeling
results.  Further supporting this notion, we note that when Eastman \& Kirshner
(1989) derived an EPM distance to SN~1987A with models custom-crafted for this
unusual event, they found $D = 49 \pm 6$ kpc, a value that is quite consistent
with the Mitchell et al. (2002) value.

In a similar manner, Baron et al. (1995)\footnote{Note that Baron et al. (1995)
and Mitchell et al. (2002) both employ the same radiative transfer code,
PHOENIX (Baron et al. 1994), whereas the Eastman et al. (1996)
values were derived using the radiative transport code EDDINGTON (Eastman \&
Pinto 1993).}  derive a SEAM distance to SN~1993J, a ``Type IIb'' event
(Filippenko 1988) thought to arise from a progenitor that exploded with just a
low-mass outer layer of hydrogen remaining since the spectrum showed hydrogen
at early times but subsequently became helium-dominated (see Matheson et
al. 2000, and references therein).  Using models customized to match the
observed spectra of SN~1993J, Baron et al. (1995) derive a distance of $4.0 \pm
0.6$ Mpc, which is in good agreement with the Cepheid distance (corrected for
metallicity effects) recently reported by Freedman et al. (2001) to its host,
M81, of $3.63 \pm 0.13$ Mpc.  Baron et al. (1995) compare the dilution factors
derived from their models of SN~1993J with those of Schmidt (1993), in which a
preliminary version of the dilution factors ultimately published by Eastman et
al. (1996) was presented.  Although it was not possible to make a detailed
comparison at the time (Schmidt only presents the dilution factors in
graphical form), Baron et al. (1995) noted that their $\zeta(T_c)$ values for
SN~1993J were generally greater than those of Schmidt (1993) by about $60\%$.
This implies that applying the EPM to SN~1993J with the dilution factors
appropriate for an SN II-P should lead to an underestimate of the distance.
Indeed, when Schmidt et al. (1993) apply the dilution factors of Schmidt
(1993) to SN~1993J, they find $D = 2.6 \pm 0.4$ Mpc, a value $\sim 50\%$ closer
than that derived by Baron et al. (1995).  As was the case with Hamuy (2001)
for SN~1987A, Schmidt et al. (1993) point out the potential danger in applying
the dilution factors crafted for SNe II-P to SN~1993J, concluding that
``supernova models that more closely match the atypical spectral features of
SN~1993J may change the inferred distance.''

Therefore, although discrepancies exist between the dilution factors presented
by Eastman et al. (1996) and those provided by one other group for SN~1987A and
SN~1993J, we conclude that at least some of the observed difference may be
explained by the fact that Eastman et al. (1996) specifically modeled SNe II-P
atmospheres, whereas the SNe modeled by the other group were both peculiar SN
II events.  Lacking a direct comparison between other modeling groups' values
for the dilution factors derived specifically to SNe II-P and those presented
by Eastman et al. (1996), it remains difficult to quantify the degree of
systematic uncertainty that the dilution factor adds to distances determined to
SNe II-P at this time; a direct comparison of the dilution factors between
modeling groups for SN II-P atmospheres is urged to help better quantify this
potential source of systematic uncertainty in the EPM technique.

For our study of SN~1999gi, by all accounts a ``normal'' SN II-P, we shall use
the dilution factors of Eastman et al. (1996) as slightly modified by Hamuy et
al. (2001), for the $BV$, $BVI$, and $VI$ filter combinations.  As recommended
by L02, we determine the photospheric velocity of SN~1999gi at each spectral
epoch by taking the weighted average of the blueshift of the absorption troughs
of the available weak, unblended line features due to \ion{Fe}{2} $\lambda
4629$, \ion{Sc}{2} $\lambda 4670$, \ion{Fe}{2} $\lambda 5276$, and \ion{Fe}{2}
$\lambda 5318$.  These features are not present in the three earliest spectra,
so we follow L02 and use the weakest available lines that are found to yield
the lowest consistent velocity at these times, namely \hdelta, \hgamma, \hbeta,
and \ion{He}{1} $\lambda 5876$.  The $BVI$ magnitudes of SN~1999gi at each of
the spectral epochs were derived through interpolation from the nearby values.
Table~4 lists the values of the parameters used in the EPM analysis.

\begin{figure}
\vskip -2.0in
\hskip -0.5in
\rotatebox{0}{
\scalebox{1.2}{
\plotone{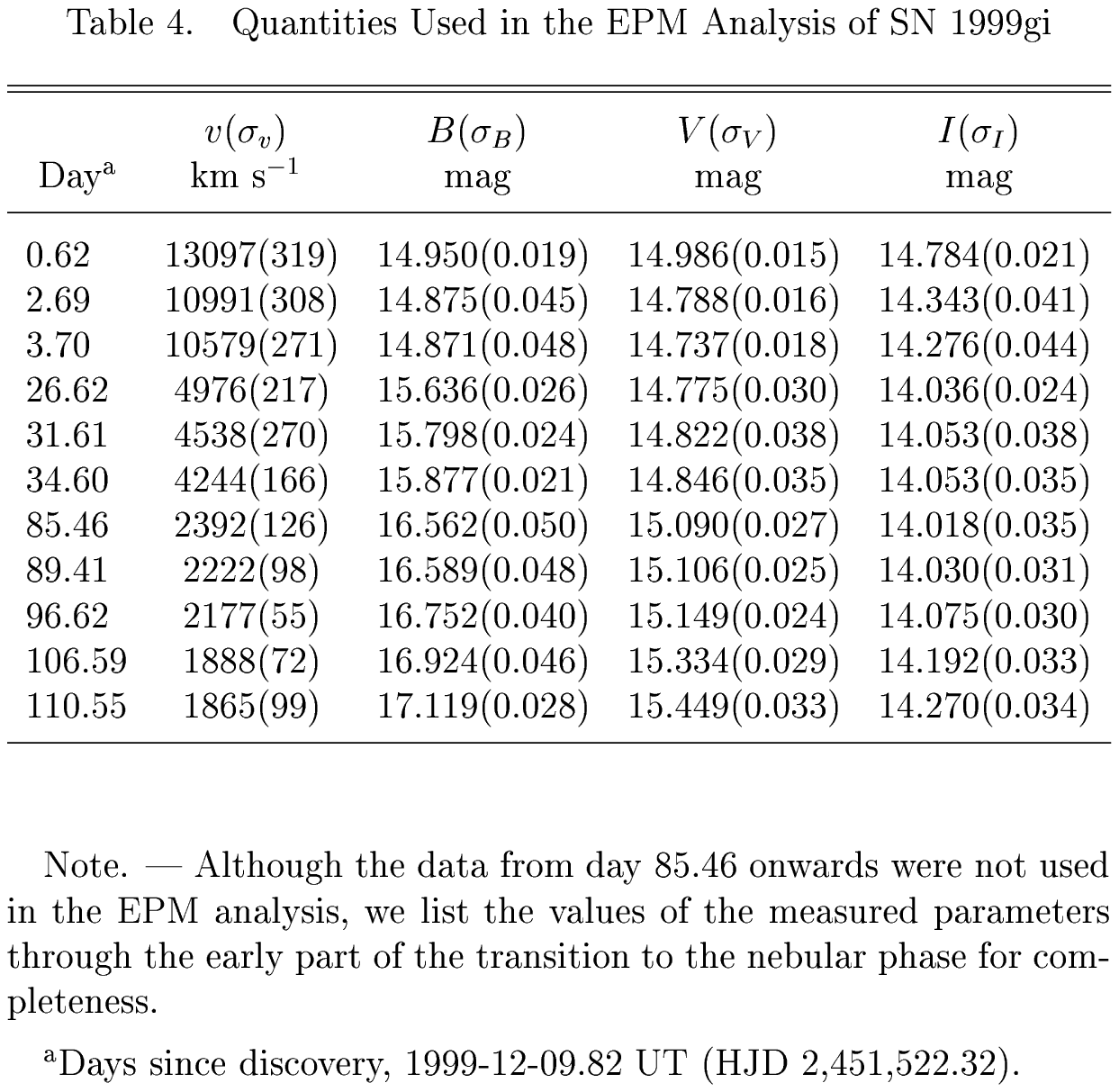}
} }
\end{figure}

Because reddening affects the inferred color temperature as well as the
apparent SN brightness, the value of $\theta$ (Eq.~\ref{eq:5}) and, therefore,
the derived distance and time of explosion will surely have a reddening
dependence.  One strength of the EPM technique is that the derived
distance should be rather robust to uncertainty in reddening since both $f$ and
$B$ in Equation~\ref{eq:5} have similar dependencies: higher reddening
decreases both the observed flux and the color temperature in a smooth, regular
manner.  However, one must also consider the temperature dependence of the
dilution factor, $\zeta$, which behaves quite differently for the various
filter combinations.  For the $BV$, $BVI$, and $VI$ filter combinations,
$\zeta$ is rather insensitive to color temperature for $T_c \gtrsim 8000$ K
(see Eastman et al. 1996; Hamuy et al. 2001). As the temperature drops below
8000 K, $\zeta$ rises for all three filter combinations, and the strength of
the temperature dependence varies considerably, from a rather weak dependence
in $VI$ to stronger dependencies in $BVI$ and, especially, $BV$.

How might we expect uncertainty in the reddening to affect the derived
distances and explosion dates for our data set?  Our spectral measurements of
SN~1999gi essentially sample two different epochs, one very early ($t < 4$ days
after discovery) and the other later during the plateau ($26 < t < 35$
days).  At the early times, the color temperatures derived from all three
filter combinations is greater than 8000 K; we therefore expect little change
in $\zeta$ for different assumed reddenings at these times.  For the later set
of data points, however, $T < 8000$ K, and $\zeta$ will be quite sensitive to
reddening changes, especially in the $BV$ and $BVI$ filter combinations.  Now,
the slope (distance) and $y$-intercept (time of explosion) of the line
described by Equation~(\ref{eq:2}) are affected somewhat differently by changes
in $\theta$.  A uniform change in the overall {\it level} of $\theta$ at each
epoch (i.e., an offset) will only affect the $y$-intercept ($t_\circ$).  A
differential change in $\theta$, on the other hand, will affect both the the
$y$-intercept and the slope ($D$).  The rapidly changing values of $\zeta$ with
$T_c$ in the $BV$ and $BVI$ filter combinations should therefore lead to a
relatively greater dependence on reddening for the derived distances in these
filter combinations than for $VI$.  On the other hand, since all filter
combinations are sensitive to the more uniform changes in $\theta$ brought
about by changes in $f$ and $B$, we might anticipate the derived explosion
times to be more equally affected by reddening uncertainty in the 3 filter
combinations.

\begin{figure}
\ssp
\begin{center}
\rotatebox{0}{
\scalebox{0.7}{
\plotone{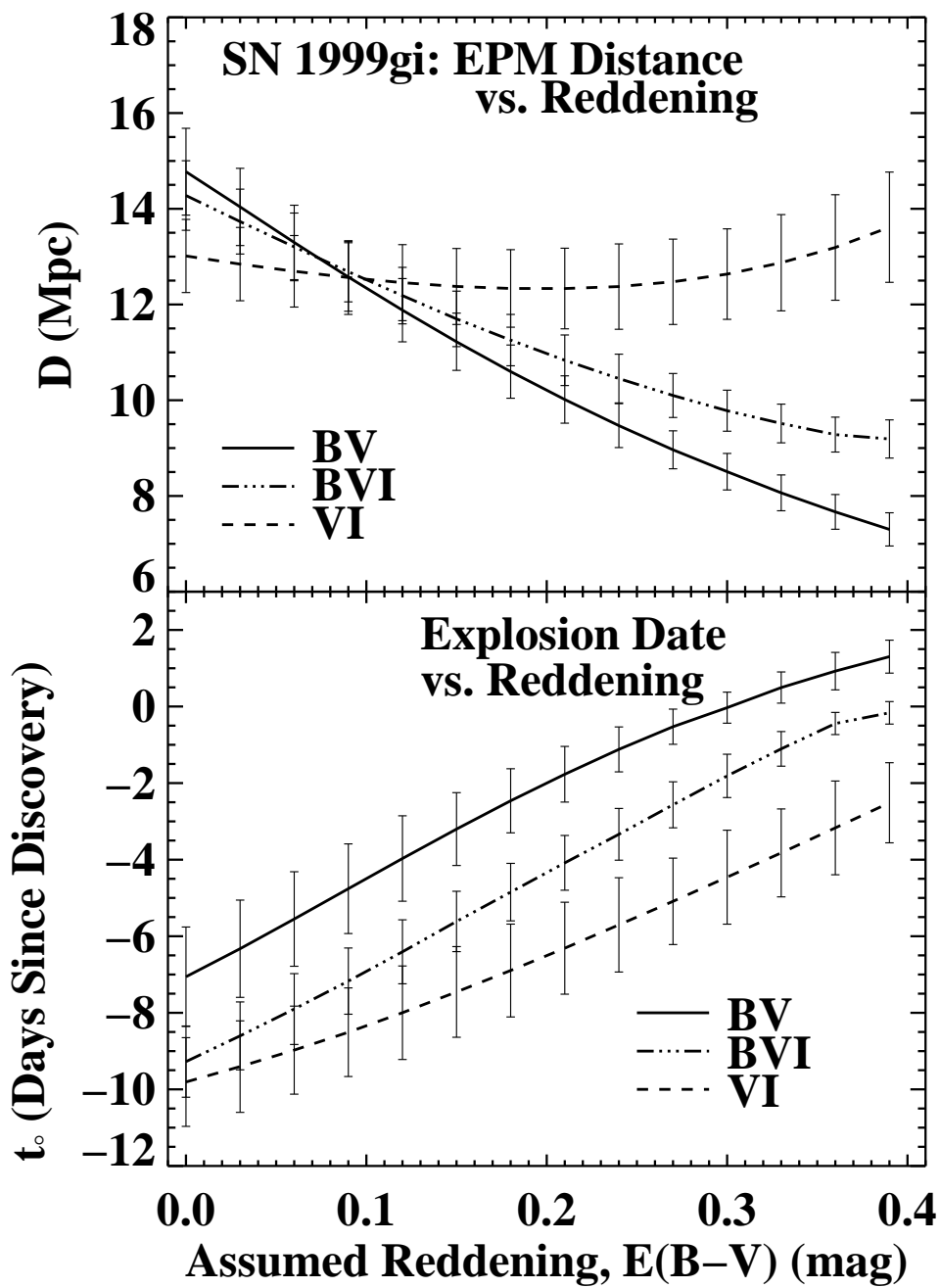}
}
}
\end{center}
\caption{Derived EPM distances ({\it top}) and explosion times ({\it bottom})
as a function of assumed reddening for SN 1999gi using three different filter
combinations to estimate the photospheric temperature and flux dilution factor.
\label{fig:8}  }
\end{figure}

The EPM distances and explosion times derived for $BV$, $BVI$, and $VI$ as a
function of assumed reddening are shown in Figure~\ref{fig:8}; to calculate the
values we have limited the EPM analysis to the first six spectral epochs (see
discussion below, and Fig.~\ref{fig:9}).  To give a sense of the formal
uncertainty in the distance for each filter combination, $1\ \sigma$
(statistical) error bars sampling reddening increments of $\Ebv = 0.03$ mag are
also shown.  These errors represent the $1\ \sigma$ spread in the distances
derived for each of the filter combinations from 1000 simulated sets of data
characterized by the values and uncertainties given in Table~4.  As
anticipated, distances derived using the $BV$ and $BVI$ filter sets are more
sensitive to changes in reddening than are distances determined using the $VI$
filter combination.  The convergence of the distances derived using the
different filter combinations at $\Ebv \approx 0.1$ mag agrees with the
conclusion of \S~\ref{sec:reddening} that SN~1999gi suffers little extinction,
and builds further confidence that the EPM technique itself may be used to
estimate SN II-P reddenings (e.g., Hamuy et al. 2001).  For the purpose of
deriving the final EPM distance and its uncertainty, we shall use the reddening
value derived in \S~\ref{sec:reddening}, $\Ebv = 0.21 \pm 0.09$ mag, which
represents the average of the EPM-derived reddening along with the 4 other
reddening estimates discussed in \S~\ref{sec:reddening}.


The value of $t_\circ$ in Figure~\ref{fig:8} also shows somewhat better mutual
consistency among the different filter combinations at lower reddenings, but
there is no strong preference for a particular reddening value.  The
inconsistency among the different filter combinations for the derived explosion
dates is troubling, and may result from the use of ``average'' dilution factors
rather than a set custom made for SN~1999gi.

We determine the final distance and time of explosion of SN~1999gi by taking
the simple means of the values derived from the three filter combinations for
$\Ebv = 0.21$ mag, and find $D = 11.06$ Mpc and $t_{\circ} = 4.06$ days before
discovery ($D = $ 10.02, 10.84, and 12.33 Mpc and $t_{\circ} = -1.79, -4.09,
{\rm and } -6.31$ days, for the $BV$, $BVI$, and $VI$
bandpasses, respectively).  Although it is difficult to assign a precise
statistical error bar to this result, there are two identifiable sources that
do contribute quantifiable uncertainty to the values of $D$ and $t_{\circ}$.
First, there is the $1\ \sigma$ spread in the values found from the three
individual filter combinations for $\Ebv = 0.21$ mag, which are $\Delta D_{\rm
filters} = 1.17$ Mpc and $\Delta t_{\rm filters} = 2.26$ days.  The second
source of uncertainty comes from the uncertainty in the reddening estimate of
$0.09$ mag.  This uncertainty leads directly to an uncertainty in distance of
$\Delta D_{\rm reddening} = ^{+1.62}_{-1.39}$ Mpc and an uncertainty in the
explosion date of $\Delta t_{\rm reddening} = ^{+1.96}_{-2.08}$ days. The final
estimate of the statistical uncertainty is found by taking the quadrature sum
of the two sources of error.  From this, then, our best estimate of the
distance and time of explosion of SN~1999gi is $D = 11.1^{+2.0}_{-1.8}$ Mpc
and $t_{\circ} = 4.1^{+3.0}_{-3.1}$ days before discovery.  Although the
uncertainty is somewhat larger than the statistical error associated with the
individual distances (especially using the $BV$ and $BVI$ filter combinations),
we feel it more accurately reflects the overall uncertainty in the distance,
since we have no {\it a priori} preference for any of the filter combinations,
and our reddening estimate suffers from uncertainty as well.

To give an idea of the relevant derived parameters, we list the dilution
factor, photospheric angular size, and photospheric color temperature for $\Ebv
= 0.21$ mag in Table~5, and plot the values used in the EPM analysis for this
reddening in Figure~\ref{fig:9}.  In Figure~\ref{fig:9} we see that the data
obtained near the very end of the plateau ($t > 85$ days; open circles) are
somewhat inconsistent with the earlier data; this is probably due to the fact
that the theoretical models of Eastman et al. (1996) include no spectra with
color temperatures as low as those derived for SN~1999gi at these late times
(Table~5), and so the inferred dilution factors rely completely on
extrapolation.  At these low temperatures, the dilution factor, especially in
$BV$ and $BVI$, is changing very rapidly with temperature.  The EPM distance is
therefore more securely obtained from the earlier epochs, and we thus have only
used the data from the first six spectral epochs to derive the distance to
SN~1999gi.

\begin{figure}
\vskip +0.in
\hskip -2.0in
\rotatebox{180}{
\scalebox{1.1}{
\plotone{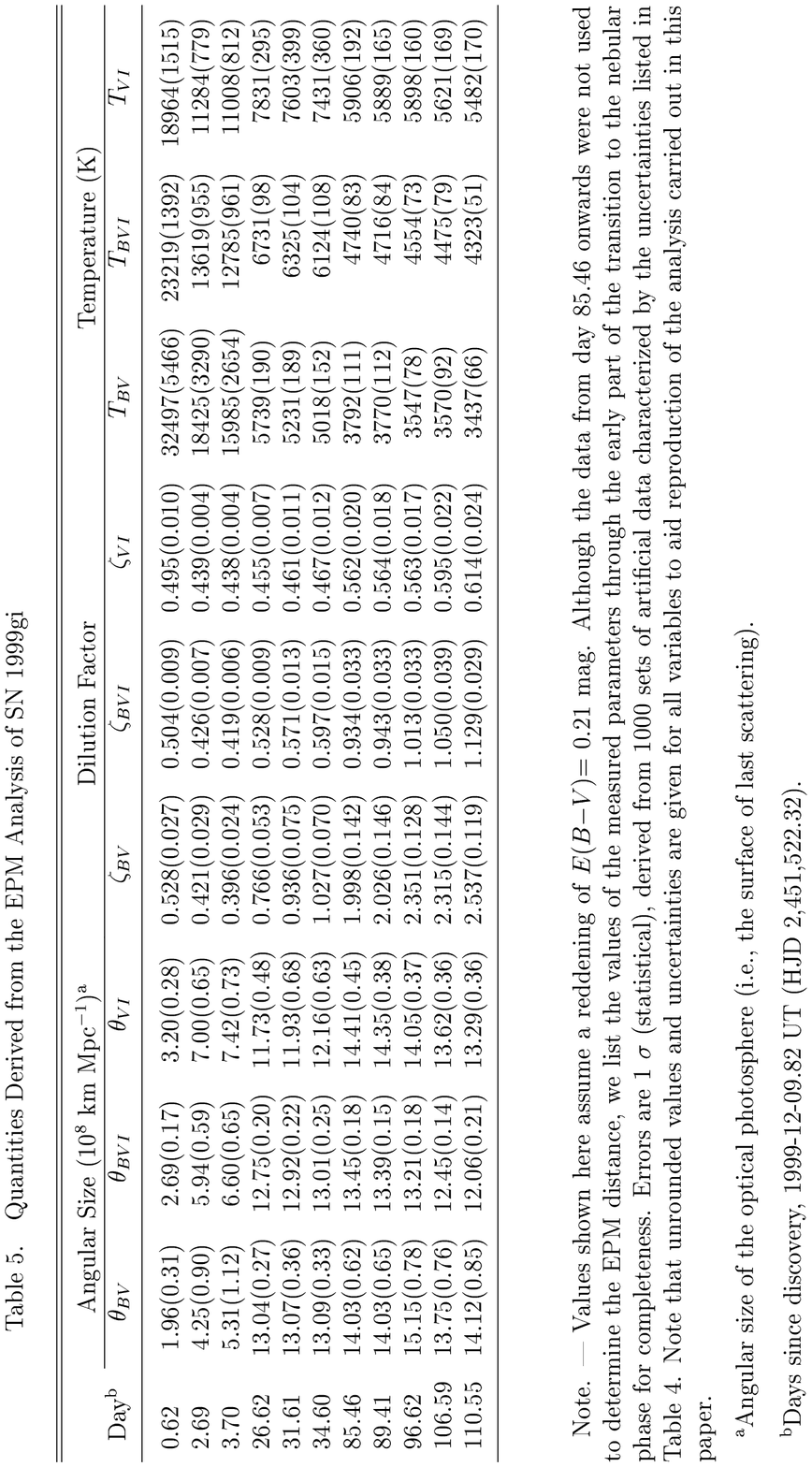}
} 
}
\end{figure}


\begin{figure}
\ssp
\begin{center}
\rotatebox{0}{
\scalebox{0.7}{
\plotone{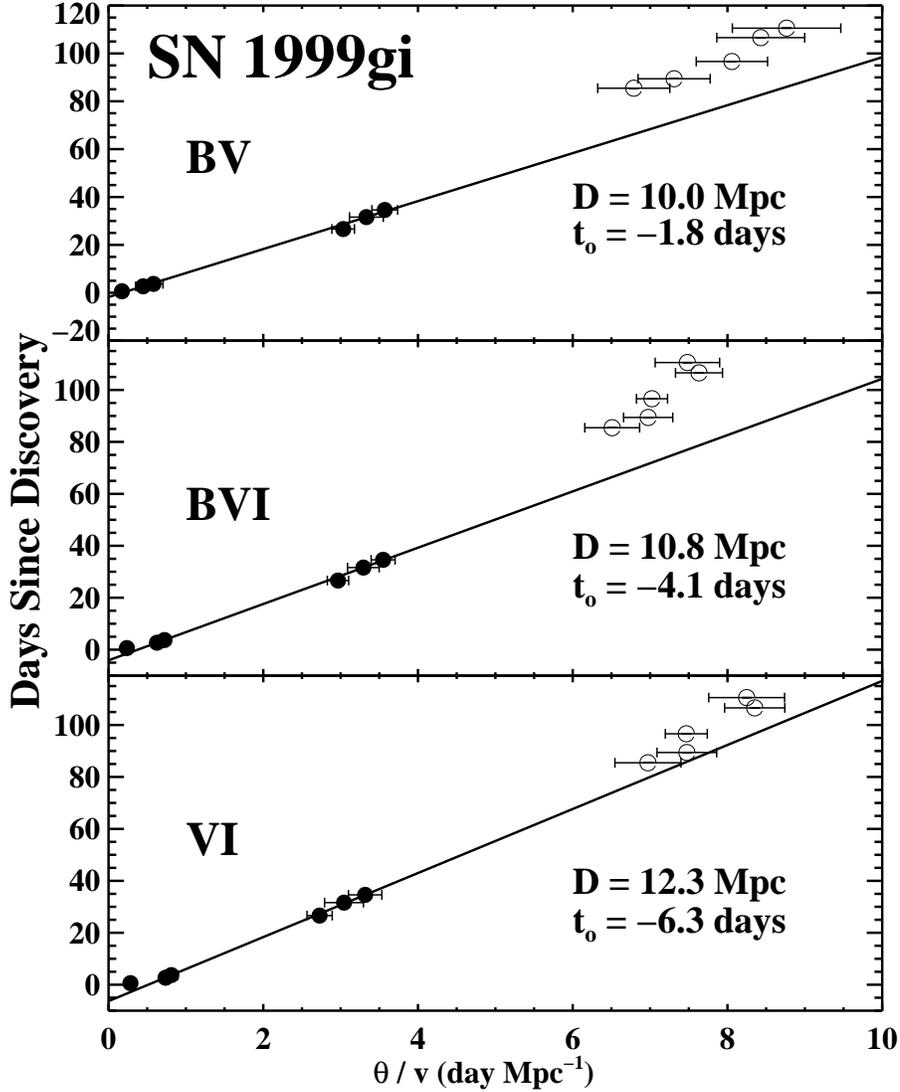}
}
}
\end{center}
\caption{The parameters used to derive the distance to SN~1999gi, with
photospheric color temperature determined using the $BV$ ({\it top}), \bvi\
({\it middle}), and $VI$ ({\it bottom}) filter combinations; a reddening of
$\Ebv = 0.21$ mag has been assumed.  $\theta$ is the angular size of the
optical photosphere (Table~5) and $v$ is the velocity of the gas at the
photosphere (Table~4).  The line of best fit is determined by applying the
criterion of least absolute deviations to the first six epochs ({\it filled
circles}).  The slope of this line yields the distance, $D$, and the
$y$-intercept the time of explosion, $t_o$.  The deviation of the later data
points ({\it open circles}) from the best-fit line may result from the lack of
theoretical models in Eastman et al. (1996) corresponding to the low
photospheric temperatures at this late photospheric phase from which to
estimate the dilution factor.
\label{fig:9}  }
\end{figure}

\section{Discussion}
\label{sec:discussion}

\subsection{EPM Distance}
\label{sec:epmdistance}

The distance to NGC~3184 has previously been estimated through a variety of
different techniques, and it is useful to compare these with our EPM distance
of $D = 11.1^{+2.0}_{-1.8}$ Mpc.  The earliest estimate is that of de
Vaucoleurs (1979), who finds $D = 7.9$ Mpc from tertiary indicators (diameters,
magnitudes, and luminosity index of spiral galaxies).  Pierce (1994) derives a
Tully-Fisher (TF) distance to NGC~3184 of $D = 7.2 \pm 1.7$ Mpc, although he
points out that the low inclination ($i = 24^{\circ}$) of NGC~3184 makes this
distance more uncertain than the statistical error bar alone suggests.  Using a
similar approach, but with updated models and parameters, Paturel et al. (in
preparation; the result is listed in the LEDA database prior to formal
publication) derive $D = 11.59$ Mpc.\footnote{This distance estimate, given by
the ``mup'' parameter in the LEDA database (i.e., using the TF method for
spiral galaxies), may change somewhat in the future when a more sophisticated
calculation is implemented, which properly accounts for the effects of
statistical bias (G. Paturel, personal communication).}  From the recession
velocity corrected for Virgo infall (Bottinelli et al. 1986; Sandage \& Tammann
1990), and assuming $H_{\circ} = 75 $ \kms\ Mpc $^{-1}$, LEDA also lists a
kinematic distance of $D = 10.09$ Mpc.  Finally, although relatively isolated,
NGC~3184 is cataloged to be in a small group of four galaxies (Tully 1988;
Giuricin et al. 2000), two of which have had Cepheid distances derived by the
{\it HST} Key Project (see Freedman et al. 2001, and references therein).  The
final Key Project distances to these two possible group-member galaxies,
corrected for metallicity, are $D = 13.30 \pm 0.55$ Mpc (NGC 3319) and $D =
13.80 \pm 0.51$ Mpc (NGC 3198).

The distance estimates to NGC~3184 (or galaxies that may be in its group)
therefore span a range $7.2 \lesssim D \lesssim 13.8$ Mpc.  Our EPM distance of
$11.1^{+2.0}_{-1.8}$ Mpc is certainly consistent with this range, and finds
best agreement with the updated TF and kinematic distances listed in LEDA.  We
note, however, that the EPM distance is somewhat inconsistent with the two
Cepheid group-member distances, in the sense $D_{\rm EPM} < D_{\rm Cepheid}$ by
$\sim 20\%$ for the average of the two Cepheid distances.  Since the
uncertainty inherent in comparing distances derived among putative group-member
galaxies could certainly account for much of the discrepancy, it is not clear
how meaningful this comparison is.  Along these lines, we point out that the
direct comparison between a Cepheid distance to a host galaxy of an SN II-P
(the type for which EPM is most securely applied) has only been carried out
thus far for one object, SN~1973R (see L02 for a complete listing of
Cepheid-EPM distance comparisons), and unfortunately the EPM distance is highly
uncertain ($\sim 50\%$).  The {\it HST} Cycle 10 program to directly compare
the Cepheid-based distance to the host galaxy of SN~1999em (D. C. Leonard et
al., in preparation) with the EPM distances derived by Hamuy et al. (2001) and
L02 will offer one solid test of the consistency of these two primary
extragalactic distance indicators.  Given the importance of firmly establishing
the value of $H_0$ from primary distance indicators, additional direct
comparisons between the Cepheid distances of the host galaxies of well-observed
SNe II-P and the EPM distances seem warranted.

Knowing the distance to SN~1999gi allows us to compare its absolute brightness
with that of previous SNe II-P.  For 9 photometrically confirmed SNe~II-P with EPM
distances, L02 find the mean plateau absolute magnitude to be
$\overline{M}_V~{\rm (plateau)} = -16.4\pm{0.6}$ mag, with SN~1999em itself
having $\overline{M}_V~{\rm (plateau)} = -15.9\pm{0.2}$ mag.  Correcting for
$A_V = 0.65$ mag (i.e., adopting $\Ebv = 0.21$ mag and $R_V = 3.1$ [Savage \&
Mathis 1979]), we measure $\overline{M}_V~{\rm (plateau)} =
-16.0 \pm 0.4$ mag for SN~1999gi, where the uncertainty incorporates both the
uncertainty in the EPM distance as well as the uncertainty in the reddening.
Again, the similarity between SN~1999gi and SN~1999em argues that these were
very similar events.  The consistency of $\overline{M}_V~{\rm (plateau)}$ for
SN~1999gi with previous SNe II-P also strengthens the suggestion by L02 that
distances good to $\sim 30\%$ ($1\ \sigma$) may be possible for SNe II-P by
simply measuring $\overline{m}_V~{\rm (plateau)}$, obviating the need for a
complete EPM analysis, unless a more accurate distance is
desired.\footnote{Hamuy \& Pinto (2002) refine this technique, showing that a
measurement of the plateau magnitude and the ejecta expansion velocity
potentially yields a considerably smaller uncertainty in the derived distance.}

\subsection{Progenitor Mass}
\label{sec:progenitormass}

Recently, S01 and S02 utilized a relatively new technique to set limits on the
progenitor masses of core-collapse SNe.  Essentially, the procedure is as
follows. (1) Search pre-explosion images of the SN's host galaxy (either
ground-based or, preferably, from {\it HST}) to look for a progenitor star at
the location of the SN (the use of post-explosion {\it HST} images is
especially helpful here to pinpoint the exact SN location). (2) If no
progenitor is found, determine the detection threshold of the image. (3) Assume
a distance to the galaxy and a reddening to the SN (and assume it is the same
as the reddening to the progenitor star) and use the appropriate bolometric
corrections to translate the detection threshold into an upper bound on the
progenitor luminosity. (4) Use \ion{H}{2} region studies of the host galaxy to
establish a likely metallicity for the progenitor star. (5) Use stellar
evolution models to predict the final pre-explosion luminosity and effective
temperature of stars spanning the range of initial masses believed to result in
core-collapse SNe of the observed type (e.g., for SNe II-P resulting from
metallicity $Z = 0.04$ stars, S01 eliminate all progenitors with $M > 25\ {\rm
M}_{\odot}$ since such stars are expected to undergo significant mass loss,
which is deemed to be inconsistent with the observed characteristics of SNe
II-P).  Finally, (6) determine which progenitor masses would have remained
undetected given the detection threshold of the image.  Since more massive
main-sequence stars generally result in brighter progenitors, it may be
possible to set an upper mass limit for an undetected progenitor star through
this approach.

The first of two objects examined by these studies is SN~1999gi (the other is
SN~1999em), for which two pre-discovery {\it HST} WFPC2 images of NGC~3184
exist, one taken through the F606W filter (central wavelength $5957$ \AA) for
160 s, and the other through the F300W filter (central wavelength $2911$ \AA)
for 800 s.  Using two post-explosion {\it HST} images to pinpoint the SN
position, S01 are unable to detect a progenitor star in either image.  From
simulations with synthetic stars, S01 derive the image detection limits as well
as the uncertainty in these limits.  To turn this into a luminosity limit, they
then assume a distance to NGC~3184 of $D = 7.9$ Mpc.  This distance is adopted
due to the general agreement that they found among three estimates: (1) the
kinematic distance using the Strauss et al. (1992) recession velocity of $592$
\kms\ and $H_0 = 75$ \kms\ Mpc$^{-1}$, with no correction for Virgo infall
(yields $D = 7.9$ Mpc); (2) the TF distance derived by Pierce (1994; $D = 7.2
\pm 1.7$ Mpc); and (3) the distance estimated by tertiary indicators (de
Vaucoleurs 1979; $D = 7.9$ Mpc).  From their analysis of the colors of the
stars in the OB association to which the progenitor of SN~1999gi presumably
belonged, S01 conclude that $\Ebv \approx 0.15$ mag, in good agreement with the
values found by our other estimates (\S~\ref{sec:reddening}).  The \ion{H}{2}
region study of NGC~3184 by Zaritsky et al. (1994) leads S01 and S02 to
conclude that the cluster containing SN~1999gi has a metallicity somewhere
between solar and twice solar, and so they adopt $Z = 0.04$ for the progenitor.
They then use the Geneva evolutionary tracks (Meynet et al. 1994; Schaller et
al. 1992), which follow stars up to the point of core carbon ignition, to
estimate the pre-explosion luminosity of progenitor stars for $7 - 40\ {\rm
M}_{\odot}$.  Comparing the progenitor star luminosities with the image
detection threshold, then, leads S01 to set an upper limit for the progenitor
of SN~1999gi of $9^{+3}_{-2}\ {\rm M}_{\odot}$.


\begin{figure}
\ssp
\begin{center}
\rotatebox{0}{
\scalebox{0.7}{
\plotone{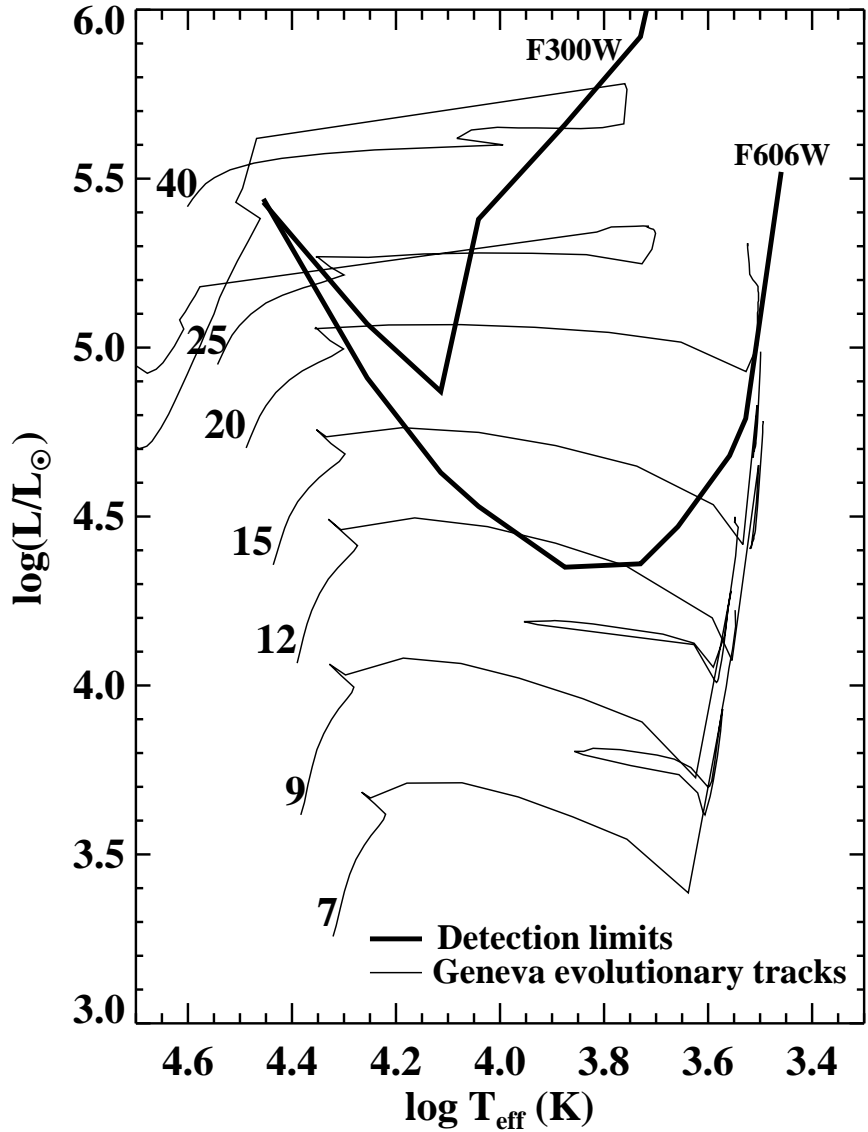}
}
}
\end{center}
\caption{The Geneva evolutionary tracks (Meynet et al. 1994; Schaller et
al. 1992) for $7-40$ M$_\odot$ stars ({\it thin lines}) plotted with the lower
detection limit of the pre-discovery image taken with the F300W and F606W
filters ({\it thick lines}) with {\it HST} as determined by S01 and modified to
reflect the EPM distance and reddening to SN~1999gi derived in this work.
Using these models, the non-detection of the progenitor star implies an upper
mass limit of $15^{+5}_{-3}\ {\rm M}_{\odot}$ for the progenitor.  
\label{fig:10}  }
\end{figure}

In a similar analysis, but for the progenitor of SN~1999em, S02 introduce an
improved stellar evolutionary model based on the most recent version of the
Eggleton (1971, 1972, 1973) evolution program, which follows stars all the way
through their carbon-burning lifetimes.  It is shown that the final
luminosities predicted for progenitor stars of $7 - 12\ {\rm M}_{\odot}$ are
significantly affected by the carbon-burning stage, curiously predicting stars
of $7 - 10\ {\rm M}_{\odot}$ to be more luminous in their final stages than
those with $11 - 15\ {\rm M}_{\odot}$.  Despite these changes in the
evolutionary code, S02 state that a reanalysis of the SN~1999gi data using the
new stellar models results in the same upper limit as that previously derived.
Since stars less massive than $\sim 8 - 10\ {\rm M}_{\odot}$ are not expected
to undergo core collapse, an upper mass limit of $9^{+3}_{-2}\ {\rm M}_{\odot}$
for the progenitor of SN~1999gi is highly restrictive, and narrows the range of
possible progenitor masses to be essentially between $8$ and $12\ {\rm
M}_{\odot}$.  Interestingly, for SN~1999em, S02 find it very difficult to
reconcile the lack of a detection with the theoretically predicted final
luminosities of the progenitor stars: Given the stated limits, {\it all} stars
with initial mass $> 7\ {\rm M}_{\odot}$ should have been detected.  It is only
by considering the {\it uncertainty} on the threshold limit ($\sim 0.2$ dex)
that S02 are able to (barely) explain the non-detection, and then only for a
very narrow range of possible progenitor masses, $12 \pm 1\ {\rm M}_{\odot}$.

We now reexamine the progenitor mass limit of SN~1999gi in light of our present
study.  We note that the distance adopted to NGC~3184 by S01 is significantly
shorter than the more recently derived distances (\S~\ref{sec:epmdistance}),
including our EPM distance to SN~1999gi of $D = 11.1^{+2.0}_{-1.8}$ Mpc
(\S~\ref{sec:sn1999giepmdistance}).  Naturally, a longer distance will result
in a brighter intrinsic luminosity detection threshold.  Reproducing the
analysis of S01, Figure~\ref{fig:10} shows the revised detection limits using
$D = 11.1$ Mpc and $\Ebv = 0.21$ mag for the progenitor of SN~1999gi, along
with the Geneva evolutionary tracks of $7 - 40\ {\rm M}_{\odot}$ stars for $Z =
0.04$ metallicity.  From this analysis, the upper mass limit for a
non-detection of the progenitor star for SN~1999gi increases to $15^{+5}_{-3}\
{\rm M}_{\odot}$; following S01, the uncertainty is simply set by the nearest
modeled progenitor masses on either side of the observed limit.

\begin{figure}
\ssp
\begin{center}
\rotatebox{0}{
\scalebox{0.7}{
\plotone{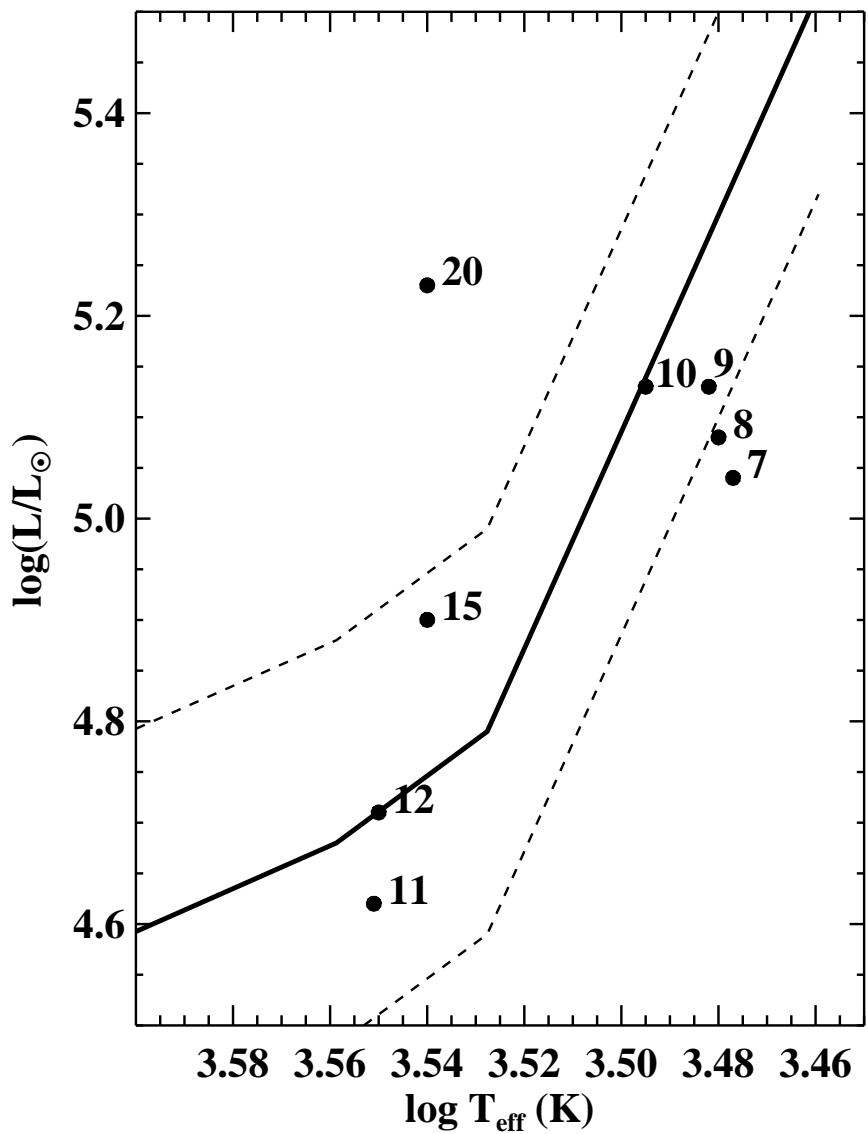}
}
}
\end{center}
\caption{Close-up of the final evolutionary state for stars of $7 - 20\ {\rm
M}_{\odot}$, using the updated models of S02.  Also shown ({\it thick solid
line}) is the detection limit of the F606W filter {\it HST} WFPC2 observation
of NGC~3184, as well as the upper and lower range of the uncertainty in the
detection limits ({\it dashed lines}).  Applying the same criteria as S02
results in an upper bound on the progenitor mass of $15^{+5}_{-3}\
{\rm M}_{\odot}$. 
\label{fig:11}  }
\end{figure}

Using the more recent stellar evolutionary models of S02 (which are only
explicitly done for $Z = 0.02$ metallicity, though metallicity changes do not
substantially alter the final pre-explosion luminosity at the low masses
considered), Figure~\ref{fig:11} shows a close-up of the expected pre-explosion
position for stars of $7 - 20\ {\rm M}_{\odot}$ along with the detection limits
of the F606W filter; the uncertainties in the detection limits (taken to be
$0.2$ dex) are also shown.  From this, we see that only progenitors with masses
greater than $\sim 12\ {\rm M}_{\odot}$ should have been detected.
Moreover, when we consider the {\it uncertainty} in the detection threshold
itself, as was done by S02, it is possible to include all progenitors $\leq 15\
{\rm M}_{\odot}$ among those that could conceivably have avoided detection.
Applying the same criterion as the most recent analysis by S02, then, results
in an updated upper mass limit for the progenitor of SN~1999gi of
$15^{+5}_{-3}\ {\rm M}_{\odot}$.


Hence, using the techniques employed by S01 and S02, but using our revised
distance and reddening estimates, we arrive at a consistent value of
$15^{+5}_{-3}\ {\rm M}_{\odot}$ as the upper mass limit for the progenitor of
SN~1999gi, compared with the upper limit of $9^{+3}_{-2}\ {\rm M}_{\odot}$
derived by S01 and S02.  The increased upper limit results mainly from the
larger distance derived through the EPM than was assumed by the earlier
analyses.  It is unfortunate that the only progenitor stars thus far positively
identified for Type II supernovae are for unusual events
(\S~\ref{sec:introduction}).  Clearly, the {\it detection} of a progenitor star
in a pre-explosion image for an SN II-P would have much to tell us about the
late stages of stellar evolution for isolated massive stars, and would serve as
a very useful check on the ability of models to accurately predict their final
luminosity.  However, since the uncertainty in the detection limits for
SN~1999gi (and SN~1999em as well) is of the same order as the theoretically
derived luminosity differences among the candidate progenitor stars, the
ability of the Smartt et al. technique to robustly discriminate among
progenitors in the range $7-15\ {\rm M}_{\odot}$ is questionable.

\section{Conclusions}
\label{sec:conclusions}

We present 15 optical spectra and 30 photometric epochs of SN~1999gi sampling
the first 169 and 174 days since discovery, respectively, and derive its EPM
distance.  Our main conclusions are as follows.

\begin{enumerate}

\item SN~1999gi is a Type II-P event with a photometric plateau lasting until
about 100 days after discovery.  It reached $B$ maximum on 1999 December $13.7
\pm 1.8$ (HJD 2,451,526.2 $\pm 1.8$, or $3.9 \pm 1.8$ days after discovery),
and achieved peak $B$ and $V$ magnitudes of $\sim 14.8$ and $\sim 14.6$,
respectively.  Overall, we find the photometric behavior of SN~1999gi to be
extremely similar to that of SN~1999em.

\item The very early-time spectra of SN~1999gi confirm the existence of the 
high-velocity absorption features in the profiles of \hbeta\ and \ion{He}{1}
$\lambda 5876$ that were first identified by Baron et al. (2000) in spectra of
SN~1999em.  The highest-velocity feature (\hbeta, day 1) extends out to nearly
$-30,000$ \kms, implying the existence of very high velocity material in
the outer envelope of SN~1999gi at early times. These features are verified to
be true P-Cygni profiles, consisting of both an absorption trough and an
emission peak in early-time spectra.  The high-velocity features are not seen,
however, in \halpha\ at early times.

\item By comparing the early-time spectral shape with blackbody functions we
derive an upper limit on the reddening of SN~1999gi of $\Ebv < 0.45$ mag;
comparison with the color evolution of SN~1999em suggests an even lower limit,
of $\Ebv < 0.30$ mag.  Other reddening estimates are consistent with these
limits, and imply a somewhat lower reddening, $\Ebv = 0.21 \pm 0.09$ mag, which
we adopt as the preferred reddening value.

\item Our best estimate for the EPM distance and explosion time of SN~1999gi is
$D = 11.1^{+2.0}_{-1.8}$ Mpc and $t_{\circ} = 4.1^{+3.0}_{-3.1}$ days prior to
discovery.  This distance is consistent with some recent distance estimates to
NGC~3184, but is $\sim 20 \%$ shorter than the average of the Cepheid distances
derived to two putative group-member galaxies.

\item The EPM distance implies an average plateau brightness of
$\overline{M}_V~{\rm (plateau)} = -16.0 \pm 0.4$ mag, which is very
similar to the value found for SN~1999em and consistent with the average
plateau brightness found by L02 of $\overline{M}_V~{\rm (plateau)} =
-16.4\pm{0.6}$ mag for 9 photometrically confirmed SNe II-P with EPM distances.

\item Following the analysis methods of S01 and S02 we derive a new upper mass
limit for the progenitor of SN~1999gi of $15^{+5}_{-3}\ {\rm M}_{\odot}$, which
is substantially less restrictive than the original limit of $9^{+3}_{-2}\ {\rm
M}_{\odot}$ found by S01 and S02.  The higher limit comes mainly from the
longer distance derived through the EPM, than was assumed by the earlier
analyses. 

\end{enumerate}

\acknowledgments

We thank Alison Coil and Maryam Modjaz for assistance with the observations,
and Georges Paturel for a helpful correspondence regarding the LEDA database
(\url{http://leda.univ-lyon1.fr}), of which we made use.  An anonymous referee
made suggestions that resulted in an improved manuscript.  We are also grateful
to the Lick Observatory and FLWO staffs for their support of the telescopes.
This research has made use of the NASA/IPAC Extragalactic Database (NED), which
is operated by the Jet Propulsion Laboratory, California Institute of
Technology, under contract with NASA.  Our work was supported in part by NASA
through the American Astronomical Society's Small Research Grant Program, and
by grants AR-8754, GO-8648, GO-9114, and GO-9155 from the Space Telescope
Science Institute, which is operated by AURA, Inc., under NASA contract NAS
5-26555.  Supernova research at the Harvard/Smithsonian Center for Astrophysics
is supported by NSF grant AST--9819825.  Additional funding was provided to
A. V. F. by NASA/Chandra grant GO-0-1001C, by NSF grant AST--9987438, by the
Guggenheim Foundation, and by the Sylvia and Jim Katzman Foundation.  KAIT was
made possible by generous donations from Sun Microsystems, Inc., the
Hewlett-Packard Company, AutoScope Corporation, Lick Observatory, the National
Science Foundation, the University of California, and the Katzman Foundation.

\end{document}